\documentclass[acmsmall,screen]{acmart}

\newcommand{\cc}[1]{\textcolor{black}{#1}} 
\newcommand{\taco}[1]{\textcolor{black}{#1}} 

\AtBeginDocument{%
  }

\setcopyright{acmlicensed}
\acmJournal{TACO}
\setcopyright{none}
\copyrightyear{2026}
\acmYear{2026}
\acmConference[]{To appear on }{2026}{}
\renewcommand\footnotetextcopyrightpermission[1]{}

\usepackage{tikz}

\usepackage{algorithm}
\usepackage{algpseudocode}
\makeatletter
\def\BState{\State\hskip-\ALG@thistlm}
\makeatother
\usepackage{amsmath}
\usepackage{multirow}
\usepackage{multicol}
\usepackage{array}
\usepackage{subfig}
\usepackage{caption}
\usepackage{balance}
\usepackage{fancyhdr}
\usepackage[absolute]{textpos}
\usepackage{listings}
\usepackage{xcolor}
\usepackage{hyphenat}
\usepackage{multirow}
\usepackage{makecell}
\usepackage{siunitx}
\usepackage{graphicx}
\usepackage{amsthm}
\usepackage{tablefootnote}
\usepackage{makecell}
\newcolumntype{?}{!{\vrule width 3pt}}

\newcolumntype{A}[1]{>{\centering\arraybackslash}p{#1}}

\newtheorem{assumption}{Assumption}
\newcolumntype{?}{!{\vrule width 1pt}}
\definecolor{codegreen}{rgb}{0,0.6,0}
\definecolor{codegray}{rgb}{0.5,0.5,0.5}
\definecolor{codepurple}{rgb}{0.58,0,0.82}
\definecolor{codeblue}{rgb}{0,0,0.92}
\definecolor{backcolour}{rgb}{0.97,0.97,0.95}

\lstdefinestyle{mystyle}{
    backgroundcolor=\color{backcolour},   
    commentstyle=\color{codegreen},
    keywordstyle=\color{blue},
    numberstyle=\tiny\color{codegray},
    stringstyle=\color{red},
    identifierstyle=\color{black},
    basicstyle=\ttfamily\scriptsize,
    emph={int,char,double,float,unsigned,void,bool,assert,std,string,vector,unique_ptr,shared_ptr,LLVM_DEBUG},
    emphstyle={\color{codepurple}},
    classoffset=1, 
    otherkeywords={>,<,.,;,-,!,=,~,*,(,)},
    morekeywords={>,<,.,;,-,!,=,~,*,(,)},
    classoffset=0,
    breakatwhitespace=false,         
    breaklines=true,                 
    captionpos=b,                    
    keepspaces=true,                 
    numbers=left,                    
    numbersep=5pt,                  
    showspaces=false,                
    showstringspaces=false,
    showtabs=false,                  
    tabsize=2,
    float=t!,
    floatplacement=t!
}
\begin{document}

\title{Protean Compiler: An Agile Framework to Drive Fine-grain Phase Ordering}



\author{Amir H. Ashouri}
\email{amirh.ashouri@huawei.com}
\orcid{0000-0001-8606-6497}
\affiliation{%
  \institution{Huawei Technologies}
  \country{Canada}
}

\author{Shayan Shirahmad Gale Bagi}
\email{shayan.shirahmad.bagi@huawei.com}
\affiliation{%
  \institution{Huawei Technologies}
  \country{Canada}
}

\author{Kavin Satheeskumar}
\email{kavin.satheeskumar1@h-partners.com}
\affiliation{%
  \institution{Huawei Technologies}
  \country{Canada}
}

\author{Tejas Srikanth}
\email{tejas.srikanth@huawei.com}
\affiliation{%
  \institution{Huawei Technologies}
  \country{Canada}
}

\author{Jonathan Zhao}
\email{jonathan.zhao1@huawei.com}
\affiliation{%
  \institution{Huawei Technologies}
  \country{Canada}
}

\author{Ibrahim Saidoun}
\email{ibrahim.saidoun@huawei.com}
\affiliation{%
  \institution{Huawei Technologies}
  \country{Canada}
}
\author{Ziwen Wang}
\email{ziwen.wang1@huawei.com}
\affiliation{%
  \institution{Huawei Technologies}
  \country{Canada}
}

\author{Bryan Chan}
\email{bryan.chan@huawei.com}
\affiliation{%
  \institution{Huawei Technologies}
  \country{Canada}
}
\author{Tomasz S. Czajkowski}
\email{tomasz.czajkowski@huawei.com}
\orcid{0009-0008-3294-6198}
\affiliation{%
  \institution{Huawei Technologies}
  \country{Canada}
}

\renewcommand{\shortauthors}{Ashouri et al.}

\begin{abstract}
 The phase ordering problem has been a long-standing challenge since the late 1970s, yet it remains an open problem due to having a vast optimization space and an unbounded nature, making it an open-ended problem without a finite solution; one can limit the scope by reducing the number and the length of optimizations. Traditionally, such locally optimized decisions are made by hand-coded algorithms tuned for a small number of benchmarks, often requiring significant effort to be retuned when the benchmark suite changes. In the past 20 years, Machine Learning has been employed to construct performance models to improve the selection and ordering of compiler optimizations; however, the approaches are not baked into the compiler seamlessly and have never been leveraged at a fine-grained scope of code segments. 

 This paper presents Protean Compiler --- an agile framework to augment LLVM with built-in phase-ordering capabilities at a fine-grained scope. The framework also comprises a complete library of more than 140 handcrafted static feature collection methods at varying scopes, and the experimental results showcase speedup gains of up to 4.1\% on average and up to 15.7\% on select CBench applications relative to LLVM's \texttt{-O3}, by just incurring a few extra seconds of build time on CBench. \taco{Additionally, Protean compiler can speed up SPEC CPU 2017 (Integer suite) by an average of 1\% and up to 6.4\% on selected benchmarks. Our framework supports parallel compilation and a built-in recommender system to speed up the search and reduce build-time overhead.}
 
 Protean compiler allows for an easy integration with third-party ML frameworks and other Large Language Models, and two applications of this two-step optimization show a gain of 10.1\% and 8.5\% speedup w.r.t. \texttt{-O3} on CBench's Susan and Jpeg applications. Protean compiler is seamlessly integrated into LLVM and can be used as a new, enhanced, full-fledged compiler. We plan to release the project to the open-source community in the near future.    
 \end{abstract}



\received{28 January 2026}
\received[revised]{May 2026}
\received[accepted]{June 2026}


\begin{textblock}{20}(1.6,0.5)
\noindent * This is a preprint version of our work, which was accepted at ACM TACO in June 2026  
\end{textblock}
\fancyhead{}

\maketitle

\section{Introduction}
\label{sec:intro}


For more than 50 years, compilers have been used not only for code generation tasks but also for code optimization \cite{Hall2009,ashouri2018survey}. The latter requires the implementation of manually tuned and handcrafted optimizations, closed and open-formed algebraic heuristics, and more recently, Machine Learning (ML) approaches to identify the profitability of an optimization pass or to identify the right set of ordering of optimization passes the pass manager needs to apply given a segment of code \cite{Kulkarni2012,Ashouri2017micomp}. \taco{The problem of phase-ordering has been an open problem in the field of compiler optimization for more than 45 years  --- A compiler pass transforms the program in ways that prohibits some optimizations that otherwise could have been performed by another subsequent compiler pass and vice-versa \cite{leverett1979overview}. This special circumstance has the potential to either cancel out the benefits of the preceding pass, i.e., the negative case, or provide the succeeding pass with enhanced opportunities as a result of its changes to the code segment. The latter is known as the positive case of phase ordering \cite{Ashouri2017micomp,ashouri2022mlgoperf,cummins2023large}.} 

Contrary to the majority of the state-of-the-art providing high-level optimizations wrapped around the compiler \cite{cummins2021compilergym,ashouri2016Cobayn,Ashouri2017micomp,poly2021-RL,wang2022automating}, recently, there have been a number of approaches integrating ML-based approaches baked into it, i.e., MLGO \cite{trofin2021mlgo}, MLGOPerf \cite{ashouri2022mlgoperf}, and ACPO \cite{ashouri2023acpo,ashouri2024work}. These leverage Ahead-Of-Time (AOT) compilation techniques to run ML models within a compiler pass of choice, e.g., Function Inlining, and to replace the profitability of its analyses by means of the inferences from the models. \taco{However, as these frameworks ML-enable single or multiple optimization passes, none have adopted a fully integrated approach to guide the pass manager to identify the best set of ordering of the phases of optimization passes.} 

Protean compiler is the first fully integrated compiler framework to augment LLVM ~\cite{lattner2004llvm} with phase-ordering of optimizations at fine-grained scopes, and it leverages an agile optimizer to find optimal solutions given a scope. Intuitively, an optimal set of ordering may be beneficial for a function or a module, but not necessarily for the whole program, and thus, we accommodate an enhanced front-end driver to take control of the visiting fine-grained scopes and, iteratively optimize them to improve performance of the generated binary. Our work utilizes an ML-based performance prediction model, i.e., IR2Score, to evaluate the intermediate transformed segments of code and to guide the agile optimizer to converge to the optimal solutions rapidly \taco{and depending on the custom scenarios, it can be a single or a pool of predictive models to be used in tandem with Protean's optimizer.} To this end, this work provides the following contributions: 

\begin{enumerate}
    \item An end-to-end framework to enable LLVM to drive phase-ordering on a fine-grained scope of code segments. These can be at the level of Module, CallGraph, Function, or Loop.
    \item A Transformer-based model, i.e., IR2Score, which receives as input the features of a scope and evaluates its predictive intermediate performance. We leverage this model to rapidly evaluate each recipe's performance, and for that, we have implemented a complete library, Protean Feature Set (PFS), of features and feature collectors at different scopes that are integrated into the compiler. Additionally, the framework allows the utilization of third-party features such as IR2VEC \cite{venkatakeerthy2020ir2vec}, and we have already integrated this project as an alternative method to showcase the flexibility of the Protean compiler. 
    \item Integration use cases of Protean framework with third-party ML-based frameworks and Large Language Models (LLMs); Not only does Protean compiler rapidly optimize standard benchmarks and user-defined code, but it can also be easily integrated with readily available frameworks and LLMs to enhance code optimization capabilities by means of phase ordering. 
\end{enumerate}

The rest of the paper is organized as follows. Section \ref{sec:related} discusses the state-of-the-art approaches for the phase-ordering problem. Section \ref{sec:proposed} gives details on our proposed framework and how the different layers of the framework communicate with each other. Subsequently, under Sections \ref{sec:results} and \ref{sec:comparison}, we showcase experimental results and comparisons with the state-of-the-art approaches. Section \ref{sec:integration} discusses integration capabilities of our work, and finally, in Section \ref{sec:conclusion}, we discuss challenges and propose directions for future work. 

\section{Related Work}
\label{sec:related}


There have been several studies to formalize and explore the optimization space \cite{Triantafyllis2003,kulkarni2009phaseOrdering,Ashouri2017micomp,ashouri2018survey}, identifying positive or negative cases of the phase-ordering problem \cite{leverett1979overview,Vegdahl1982,Ashouri2017micomp}, and reducing the dimensionality of the problem \cite{kulkarni2010improving,kulkarni2006exhaustive,cooper2002compilation,georgiou2018less,theodoridis2022understanding,Purini2013}. Nonetheless, the inherent challenge is that there is no ideal order, as the problem is unbounded, and even the bounded problem suffers from NP-hardness complexity \cite{ashouri2018survey,leverett1979overview,NobreRicardoLusReis2016}.
More recently, there have been several approaches in using ML techniques to alleviate the complexity of the problem \cite{zhao2025leveraging,haj2020neurovectorizer,huang2019autophase,mammadli2020-phaseordering,Ashouri2016predictiveModeling,Ashouri2017micomp,quetschlich2023compiler}; however, none have integrated the end-to-end methodology into a state-of-the-art compiler and mostly build upon external/third-party tools or applications to complete the end-to-end pipeline. Additionally, one of the difficulties in leveraging ML models to tackle the phase-ordering problem is the lack of labeled data in the compilation domain. Different architectures and compilers require specific tuning to maximize performance, and thus, scarce data also suffers from a lack of reusability and reproducibility \cite{fursin2021collective,fursin2025framing,fursin2024enabling}. For these reasons, several works tackle the problem of phase ordering using other metrics, e.g., code size reduction \cite{cummins2021compilergym,cummins2023large,jayatilaka2021towards}.

Cummins et al. \cite{cummins2021compilergym} propose \textit{CompilerGym}; a compiler optimization environment leveraging OpenAI's Gym, later superseded with Gymnasium \cite{1606.01540OpenAIGym}, to provide a wrapper around compiler for doing phase-ordering jobs. The authors showcase their capabilities by inferring code size optimization scenarios and Reinforcement Learning (RL) techniques. Similarly, our work also enables optimizing segments of code at a finer granularity; however, our work integrates the end-to-end phase ordering optimization approach into LLVM directly, and our primary objective is optimizing the performance of the running application rather than optimizing code size.

Ashouri et al. \cite{Ashouri2017micomp} propose MiCOMP; a phase ordering methodology that constructs subsequences of compiler optimizations and employs a predictive model to derive an iterative compilation on LLVM. This is the closest work to us; however, there are a number of stark contrasts between the two. Similar to CompilerGym, MiCOMP is also a Pythonic wrapper to enable phase-ordering explorations and optimization. Second, MiCOMP solely focuses on program-wide scope and doesn't allow finer-granularity, i.e., Module, Function, and Loops. Finally, the work leverages dynamic features in its predictive model, making it inherently slower compared with static features and pre-trained embeddings utilized by our work at inference. This is especially crucial in industrial test cases, which might take minutes to run, and leveraging dynamic features would make it infeasible to deploy. In our work, Protean's agile compilation relies on rapidly generated features and can also integrate open-source libraries such as \textit{IR2VEC} \cite{venkatakeerthy2020ir2vec} for training and deployment. 

\taco{Protean compiler is unique in the sense that, contrary to MLGO, MLGOPerf, and ACPO, which they ML-enable single or multiple optimization passes, none have adopted a fully integrated approach to guide the pass manager to identify the best set of
ordering of the phases of optimization passes. Additionally, our novelty lies in the combination of finer granularity and a tailored phase ordering recipe given a segment of a code, and we provide details of our framework in the following section.}

\section{Proposed Framework}
\label{sec:proposed}

\begin{figure}[!t]
\centering
\includegraphics[width=.65\textwidth]{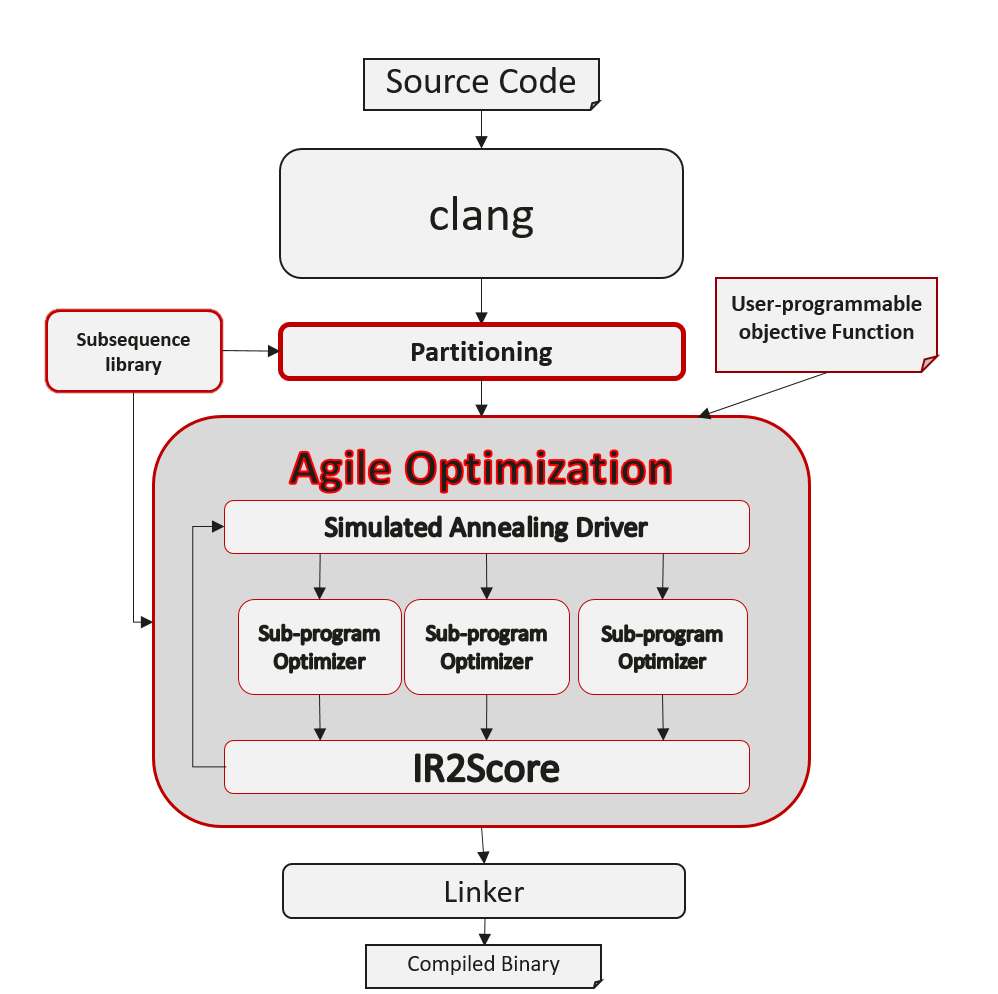}
\caption{Protean Compiler System Architecture} 
\label{fig:protean_system_diagram}
\end{figure}

The high-level architecture of our system is presented in Figure~\ref{fig:protean_system_diagram}. The flow takes as input an application source code and a specification of the optimization objective (for example, performance or code size). The process begins with {\it clang} where the input program is converted into intermediate representation, LLVM IR. The intermediate representation of the complete program is then partitioned into sub-components (or sub-programs) for parallel processing. The simplest form of such partitioning is on a per-module (per-file) basis; however, more comprehensive methods can be applied to allow for more effective grouping of code to accelerate the optimization process. Once partitions are created, the agile optimization at the core of our framework is engaged to perform iterative optimization of individual partitions, before ultimately combining them into a single binary in the linking stage.

The \textit{agile optimization} in our work is an iterative framework for recipe-based code optimization driven by a Simulated Annealing~\cite{kirkpatrick1983optimization} engine, where decisions are evaluated by an ML model we call \textit{IR2Score}. Each iteration of the process creates a pass recipe based on available pass subsequences (from the subsequences library) and then applies the recipe to a given partition. The resulting LLVM IR is evaluated by IR2Score model to produce an assessment of the quality of the solution. When the framework is used for performance optimization, the score models the amount of speedup, and thus an increase in the score indicates a predicted improvement in performance. \taco{In the case of code size reduction, IR2Score can model the reduction in the binary size or the number of instructions, and so a decrease in the value would be the objective function.} Once the IR2Score result is produced, the configuration is accepted according to its relation to the current state and the following cost function:
\begin{equation}
\left\{
\begin{array}{ll}
1 & \text{if } C_{new} > C_{present} \\
e^{-\frac{C_{new} - C_{present}}{T}} & \text{if } C_{new} \leq C_{present}
\end{array}
\right.
\end{equation}
where $C_{new}$ is the cost produced by IR2Score for the newly tested recipe, $C_{present}$ represents the cost of the present state recipe, and $T$ is the temperature setting.

This strategy is dependent on the temperature setting $T$ for the simulated annealing process, where a high temperature indicates a solution very much in flux (early simulated annealing steps), allowing for poor results to be accepted as a means of hill-climbing to allow for a broader search space exploration. Over time, we decrease the temperature, which causes the simulated annealing process to accept fewer and fewer poorer solutions, and eventually only accept better solutions. As a result, the simulated annealing process effectively searches the solution space quickly and refines its output as the temperature is lowered.
At the end of the annealing process, the Protean compiler converges on an optimized solution for each partition, where each partition may have a different recipe applied to it. The flow then combines the optimized partitions in the linker stage for the final optimization step (if LTO is used) and binary generation. \taco{When LTO is disabled, Protean allows for a native linking procedure in which the generated modules are
native machine code (.o files) as opposed to LLVM bitcode, no other optimization is applied program-wide, and only symbols are resolved by the linker to produce a binary/executable.}

In the following sections, we describe additional implementation details as they pertain to Protean compiler implementation in the LLVM~\cite{lattner2004llvm} framework, the means by which subsequences have been created, and the details of the IR2Score model used for per-partition performance assessment. Finally, we discuss how additional optimization using third-party projects or LLMs could be applied prior to agile compilation to further optimize workload performance.

\subsection{Agile Driver}
\label{sec:proposed:agile} 

To implement the Protean compiler, we used LLVM Compiler Infrastructure~\cite{lattner2004llvm} and adapted its flow to support agile compilation. To do this, we added an additional optimization level to {\it clang} to allow us to invoke the new flow simply by replacing \texttt{-O2} or \texttt{-O3} with our new Protean Optimization level (\texttt{-OP}) flag. This necessitates changes to the Clang driver and the introduction of a new compilation executable to implement the agile compilation loop.

\subsubsection{Clang Driver Modifications}

To successfully introduce a new compilation stage, the {\it clang} driver requires that the sequence of compilation {\it phases} be updated by defining a new phase, which we called {\it ProteanOpt}. We inserted our step after {\it phases::Compile} and {\it phases::Backend} steps, which in principle forces the use of the ProteanOpt step under all circumstances. However, by adding a clang flag \texttt{-OP}, we are able to filter out phase steps at runtime and skip the Protean Optimization step if the \texttt{-OP} flag is not present. Thus, we can now have a compiler where \texttt{-OP} coexists seamlessly with existing optimization flows.

With a new compilation phase created in the driver, we proceeded to add a new optimization tool called {\it protean} that implements the agile compilation flow. This tool is essentially a modified version of {\it opt} that has the same dependencies, but the main optimization step that usually invokes a fixed pipeline is now replaced with an iterative agile optimization loop. In this loop, we spawn a subprocess to apply a recipe on a module and wait for the subprocess to complete before we evaluate the resulting IR using the IR2Score model.
Note that the child process may fail abnormally due to a pass sequence that resulted in an unexpected corner case not handled by the compiler; however, since it is a subprocess, we only need to handle the abnormal exit by automatically rejecting the solution from the search space. Thus, even if the recipe triggers an unexpected corner case, the Protean compiler will simply continue while being aware of the failed attempt.

\subsection{Supporting Parallel Builds}
\label{sec:proposed:parallel}

\taco{Protean compiler supports the parallel building of a large program by ensuring that multiple module-level explorations can occur at the same time and are protected from each other. A parallel build tool, such as GNU make or Ninja, typically launches one instance of Clang per module, and at the end invokes the linker to link all the object files together to form the final output. When this setup is used with Protean, each instance of Clang dumps the unoptimized LLVM IR for a module, and the Clang driver passes the LLVM IR, along with its own process ID, to our protean tool, which runs in a subprocess to apply a recipe to the module. We allocate a \textit{shared memory buffer}, identified by the parent process ID, through which both of our integrated feature collection techniques, i.e., PFS and IR2VEC, can communicate with the protean tool. Thanks to this mechanism, each subprocess will read and write from its own buffer, without interfering with other concurrent instances of Protean compiler, and no further disk I/O operations are needed to collect the features of a module, which helps reduce the compilation overhead. Once the features are collected, they are passed to IR2Score for evaluation, and the subprocess returns once completed.}

\subsection{Protean Recommender System}
\label{sec:proposed:recsys} 

\taco{The Simulated Annealing (SA) process effectively searches the solution space and refines its output as the temperature is lowered. However, there are a number of drawbacks in its pure form, including but not limited to being a single-path focus metaheuristic, the probability of accepting local minima/maxima, and being overly correlated with the temperature/cooling parameter. For these reasons, we add a rapid and intuitive Recommender System \cite{ricci2011introduction,Ashouri2017micomp} to SA to speed up its exploration -- employing \textit{genetic algorithm} (GA) \cite{katoch2021reviewgenetic}. At a high level, GA starts with an initial population that has already been evaluated by the fitness function, i.e., IR2Score. The best recipes are selected as parents for the next generation, and they are used to produce offspring through Crossover (recombination) and Mutation (random/small changes) until the best new offspring are produced, and the algorithm is terminated.
SA, together with GA, form hybrid algorithms (frequently called GSA or SAGA) to balance global search and local refinement. While GA excels at exploring a broad search space, it can struggle with hill-climbing -- SA compensates for this by providing a robust mechanism for local optimization. Our SGA has a few hyperparameters, i.e., initial population, crossover, and mutation rate, which can play a role in tuning the exploration policy, and we mention the values later under Section \ref{sec:results}.} 

\subsection{Protean Subsequences}
\label{sec:proposed:subsequences}

\cc{A number of recent works suggest clustering the standard optimization passes, e.g., LLVM's \texttt{-O3}, into shorter subsequences \cite{Ashouri2017micomp,lemerre2023ssa,Martins2014,NobreRicardoLusReis2016, Ashouri2016predictiveModeling}}. The reason is twofold. First, the optimization space can be reduced from the full factorial set to a reduced factorial set where each string of optimization is a permutation of the combination in the form $\alpha_{i_1} \alpha_{i_2} \cdots \alpha_{i_k}$, where $i_1, i_2, \ldots, i_k \in \{ 1, \ldots, M \}$. The integer $k$ is the {\it length} of the optimization space. Second, the internally fixed ordered optimizations under each subsequence are meant to be followed by one another, as designed tentatively by the compiler developers throughout the generations of LLVM.  In reality, clustering optimizations into subsequences makes sense as certain analysis and transformation passes should be performed prior to another optimization in order for the optimization to have the optimal effect. For instance. \textit{Function Inlining} prior to \textit{Loop Unrolling} and \textit{Vectorization} can provide new opportunities which otherwise would not have been known to these optimizations \cite{ashouri2022mlgoperf}. This is also known as the positive phase ordering effect we discussed earlier in Section \ref{sec:related}. 

\subsubsection{Optimization Clustering}
\label{sec:proposed:subsequences:clustering}

\cc{There are more than 160 passes in LLVM's standard optimization level \texttt{-O3}, some of which are \textit{Analysis} passes, which provide necessary information for \textit{Transformation} passes. Inspecting the sequence more closely, we can observe that there are pockets of repetition of certain groups of passes followed by one another. As shown by earlier works \cite{Ashouri2017micomp,Martins2014,Ashouri2016predictiveModeling}, choosing the right number of subsequences in a recipe can significantly alter the optimization space comprised by the combination of permutations of these subsequences. We cluster our optimization passes by leveraging a clustering method, i.e., Agglomerative clustering \cite{mullner2011modern}. \taco{To prepare the input for our clustering algorithm, we create an optimization dependency graph of the compiler optimization sequence at \texttt{-O3} for which the nodes represent the optimizations and the edges represent the number of times pairs of optimizations appear consecutively in the \texttt{-O3} sequence. Naturally, if a pair has a higher number, it shows a stronger co-positional ordering for the pair. Once the graph is created, we can translate it into a weighted matrix, which can be fed to our clustering technique}. For brevity, we exclude details on the clustering mechanism itself, but at a high-level it does the following. (1) it is a bottom-up hierarchical unsupervised clustering method that takes as input all the optimization passes in \texttt{-O3}, (2) it calculates the weighted matrix and builds the K-nearest neighbours, (3) it builds the transition probability and forms sample clusters $\{c_1, \ldots, c_n\}$ and finally, (5) iteratively, tries to add more subcluster to the formed list by maximizing its objective function. Once the algorithm finishes, it outputs our clusters of subsequences of optimizations. Table \ref{tab:subsequences} showcases our Protean subsequences, clustered into 5 recipes, A through E. }

\begin{table*}[!t]
\small
\begin{tabular}{|l|m{13cm}|}
\hline
\textbf{A} & {\color[HTML]{000000} globalopt,\textbf{cgscc}(devirt\textless{}4\textgreater{}(inline\textless{}only-mandatory\textgreater{},inline,move-auto-init,function-attrs\textless{}skip-non-recursive-function-attrs\textgreater{},argpromotion,\textbf{function}\textless{}eager-inv;no-rerun\textgreater{}(sroa\textless{}modify-cfg\textgreater{},speculative-execution,tailcallelim,loop-mssa(licm\textless{}allowspeculation\textgreater{},simple-loop-unswitch\textless{}nontrivial;trivial\textgreater{}),\textbf{loop}(loop-idiom,indvars,loop-deletion),loop-unroll\textless{}O3\textgreater{},early-cse\textless{}\textgreater{},callsite-splitting,sroa\textless{}modify-cfg\textgreater{},early-cse\textless{}memssa\textgreater{},speculative-execution,jump-threading,correlated-propagation,lower-expect,simplifycfg\textless{}bonus-inst-threshold=1;no-forward-switch-cond;no-switch-range-to-icmp;no-switch-to-lookup;keep-loops;no-hoist-common-insts;no-sink-common-insts;speculate-blocks;simplify-cond-branch\textgreater{},instcombine\textless{}max-iterations=1;no-use-loop-info;no-verify-fixpoint\textgreater{},aggressive-instcombine,tailcallelim,simplifycfg\textless{}bonus-inst-threshold=1;no-forward-switch-cond;no-switch-range-to-icmp;no-switch-to-lookup;keep-loops;no-hoist-common-insts;no-sink-common-insts;speculate-blocks;simplify-cond-branch\textgreater{},reassociate))),)}                                  \\ \hline
\textbf{B} & {\color[HTML]{000000} \textbf{function}\textless{}eager-inv\textgreater{}(loop-simplify,lcssa,crypto,chr,\textbf{loop}(loop-rotate\textless{}no-header-duplication;no-prepare-for-lto\textgreater{},loop-deletion),annotation-remarks,constraint-elimination,mem2reg,instcombine\textless{}max-iterations=1;no-use-loop-info;no-verify-fixpoint\textgreater{},loop-simplify,lcssa,indvars,loop-deletion,loop-simplify,lcssa,loop-instsimplify,loop-simplifycfg,\textbf{function}(loop-mssa(licm\textless{}allowspeculation\textgreater{})),simple-loop-unswitch,simplifycfg\textless{}bonus-inst-threshold=1;no-forward-switch-cond;no-switch-range-to-icmp;no-switch-to-lookup;keep-loops;no-hoist-common-insts;no-sink-common-insts;speculate-blocks;simplify-cond-branch\textgreater{},instcombine\textless{}max-iterations=1;no-use-loop-info;no-verify-fixpoint\textgreater{}),require\textless{}globals-aa\textgreater{},function(invalidate\textless{}aa\textgreater{}),require\textless{}profile-summary\textgreater{},\textbf{function}\textless{}eager-inv\textgreater{}(loop-simplify,lcssa,loop(loop-idiom,loop-deletion,loop-unroll-full),loop-data-prefetch,hash-data-prefetch,separate-const-offset-from-gep),)}                                                                                                                                                                                                                                \\ \hline
\textbf{C} & {\color[HTML]{000000} \textbf{function}\textless{}eager-inv\textgreater{}(sroa\textless{}modify-cfg\textgreater{},gvn-hoist,mldst-motion,gvn,sccp,bdce,instcombine\textless{}max-iterations=1;no-use-loop-info;no-verify-fixpoint\textgreater{},jump-threading,correlated-propagation,adce,memcpyopt),)}                                                                                                                                                                                                                                                                                                                                                                                                                                                                                                                                                                                                                                                                                                                                                                                                                                                                                                                                                                                                                                                                                                                            \\ \hline
\textbf{D} & {\color[HTML]{000000} \textbf{cgscc}(dse,function\textless{}eager-inv\textgreater{}(loop-simplify,lcssa,coro-elide,simplifycfg\textless{}bonus-inst-threshold=1;no-forward-switch-cond;no-switch-range-to-icmp;no-switch-to-lookup;keep-loops;no-hoist-common-insts;no-sink-common-insts;speculate-blocks;simplify-cond-branch\textgreater{},instcombine\textless{}max-iterations=1;no-use-loop-info;no-verify-fixpoint\textgreater{},reassociate),function-attrs,\textbf{function}(require\textless{}should-not-run-function-passes\textgreater{}),coro-split,function(invalidate\textless{}all\textgreater{})),deadargelim,coro-cleanup,globalopt,globaldce,elim-avail-extern,rpo-function-attrs,recompute-globalsaa,ipsccp,\textbf{function}\textless{}eager-inv\textgreater{}(float2int,lower-constant-intrinsics),constmerge,cg-profile,rel-lookup-table-converter,ir-library-injection,)}                                                                                                                                                                                                                                                                                                                                                                                                                                                                                                                                                       \\ \hline
\textbf{E} & {\color[HTML]{000000} \textbf{function}\textless{}eager-inv\textgreater{}(loop-simplify,lcssa,\textbf{loop}(loop-rotate\textless{}no-header-duplication;no-prepare-for-lto\textgreater{},loop-deletion),loop-distribute,loop-simplify,lcssa,loop-unroll-and-jam,inject-tli-mappings,loop-vectorize\textless{}no-interleave-forced-only;vectorize-forced-only;\textgreater{},infer-alignment,loop-load-elim,instcombine\textless{}max-iterations=1;no-use-loop-info;no-verify-fixpoint\textgreater{},simplifycfg\textless{}bonus-inst-threshold=1;no-forward-switch-cond;no-switch-range-to-icmp;no-switch-to-lookup;keep-loops;no-hoist-common-insts;no-sink-common-insts;speculate-blocks;simplify-cond-branch\textgreater{},vector-combine,instcombine\textless{}max-iterations=1;no-use-loop-info;no-verify-fixpoint\textgreater{},loop-unroll\textless{}O3\textgreater{},transform-warning,sroa\textless{}preserve-cfg\textgreater{},instcombine\textless{}max-iterations=1;no-use-loop-info;no-verify-fixpoint\textgreater{},loop-simplify,lcssa,loop-mssa(licm\textless{}allowspeculation\textgreater{}),alignment-from-assumptions,loop-sink,instsimplify,div-rem-pairs,tailcallelim,simplifycfg\textless{}bonus-inst-threshold=1;no-forward-switch-cond;no-switch-range-to-icmp;no-switch-to-lookup;keep-loops;no-hoist-common-insts;no-sink-common-insts;speculate-blocks;simplify-cond-branch\textgreater{},annotation-remarks),)} \\ \hline
\end{tabular}
\caption{List of Protean Compiler Subsequences}
\label{tab:subsequences}
\end{table*}

\subsubsection{Optimization Recipe Length}
\label{sec:proposed:subsequences:maxLength}

Once we have identified our 5 clusters, a.k.a subsequences of optimizations, we need to select the max allowed length of subsequences in a recipe. We experiment with allowing the recipes grow up between a max lengths of 3 to 7 subsequences, and we observe that the optimization space comprises 156, 781, 3096, 19k, and 97k different permutations of combinations of optimizations, respectively \footnote{It is computed as the summation of $n^i$ with $i \in \{0, \ldots, m\}$ where n is the number of optimizations and m is the max allowed length of the optimization vector\label{fn:exhaustive}}. 
\taco{To identify the max subsequence length, we choose one application from each CBench category as a representative set and run an iterative compilation strategy to observe the speedup gains by varying the max length parameter. The applications are as follows: automotive\_susan\_c, consumer\_jpeg\_d, network\_dijkstra, office\_rsynth, security\_blowfish\_d, and telecom\_adpcm\_c. Table \ref{tab:subseq_len} shows the experimental results we obtained by means of a geometric mean of speedup gained. Among the set, we choose 5 because it yields orders-of-magnitude reduction in the phase-ordering optimization space relative to the baseline (length = 7), while reaching 97\% of the achievable speedup.}

\begin{table}[!t]
    \centering
    \begin{tabular}{ |>{\centering\arraybackslash}m{2.5cm}| >{\centering\arraybackslash}m{3cm}| >{\centering\arraybackslash}m{5cm}|}
    \hline
        Max Allowed Recipe Length  & Compiler Optimization Space & Relative Performance Gained by Exhaustive Iterative Compilation  \\ \hline
        3 & 156 & 0.85  \\ \hline
        4 & 781 & 0.92  \\ \hline
        \textbf{5} & 4k & \textbf{0.97}  \\ \hline
        6 & 19k & 0.99  \\ \hline
        7 (Baseline) & 97k & 1  \\ \hline
    \end{tabular}
    \caption{Analysis of the Effect of the Number of Subsequences in a Recipe (Recipe Length) on Performance}
    \label{tab:subseq_len}
\end{table}

\subsection{IR2Score Model}
\label{sec:proposed:ir2score}

\begin{figure}[!t]
\centering
\includegraphics[width=.75\textwidth]{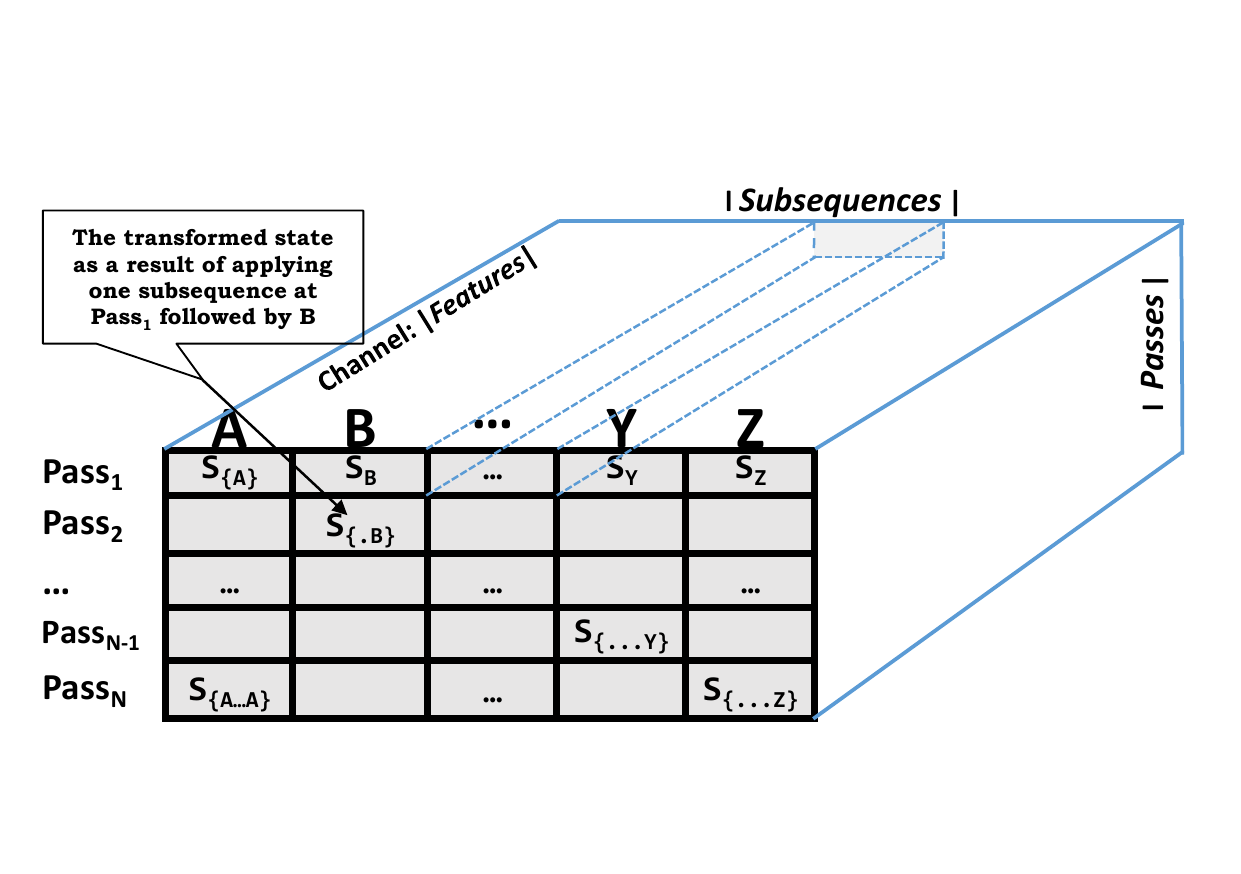}
\caption{IR2Score Model Data Representation} 
\label{fig:ir2score_cnn}
\end{figure}

As discussed earlier under Section \ref{sec:proposed}, IR2Score receives as input the features of the IR and produces as output the predicted speedup as the Cost (score) for the given sub-program.  
\begin{assumption}
\label{assum_det}
    Let's denote the transformation applied to code as a result of optimization pass $x$ as $f_x$. The result of optimizing source code $S$ using a set of optimization passes $o=\{a, b, c, d, e, ...\}$ at once, is equivalent to applying the optimization passes one by one to the optimized code as a result of the previous optimization pass, such that:
    \begin{equation}
        S_o = f_o(S) \approx f_{o[n]}(f_{o[n-1]}(... f_{o[0]}(x)))
    \end{equation}
    where $o[i]$ denotes optimization pass at order $i$.
\end{assumption}
According to Assumption~\ref{assum_det}, we can say that there exists a relation between features of a code optimized using all passes in $o$ and features of a code optimized using passes in earlier orders of $o$. To leverage this relation, we represent the input data in our IR2Score model as illustrated in Figure~\ref{fig:ir2score_cnn}. We use a Transformer model structure to learn the sequential relations discussed in Assumption~\ref{assum_det}, as it provides an intuitive architecture for the given problem. This gives us a higher accuracy and lower validation loss values compared with an MLP-like shape (fully-connected Neural Net (NN)), a Convolutional Neural Net (CNN), or an LSTM-like shape, as it leverages the feature map and preserves the history of the current IR by keeping its previous state's features in the learning process \taco{and we use this proxy for choosing the design. IR2Score receives as input the features from the LLVM IR at the current state and outputs the predicted speedup with respect to the baseline, i.e., -O3. Because of its Transformer-like design, it can also receive sequential feature states together with the current state as well, but this would be optional and, upon availability. As shown at Footnote \ref{fn:exhaustive}, the phase ordering space becomes enormously large when we take into account the number of functions-module-application-recipe pairs by means of an exhaustive permutation; therefore, it shows how challenging it is to generate quality labeled data for performance optimization purposes.}

\subsection{Protean Feature Set}
\label{sec:proposed:proteanft}

\begin{figure}[!t]
\centering
\subfloat[Protean Feature Set (PFS) Pearson Correlation]{
\label{fig:ProteanFTCorr}
\resizebox{.44\textwidth}{!}{\includegraphics{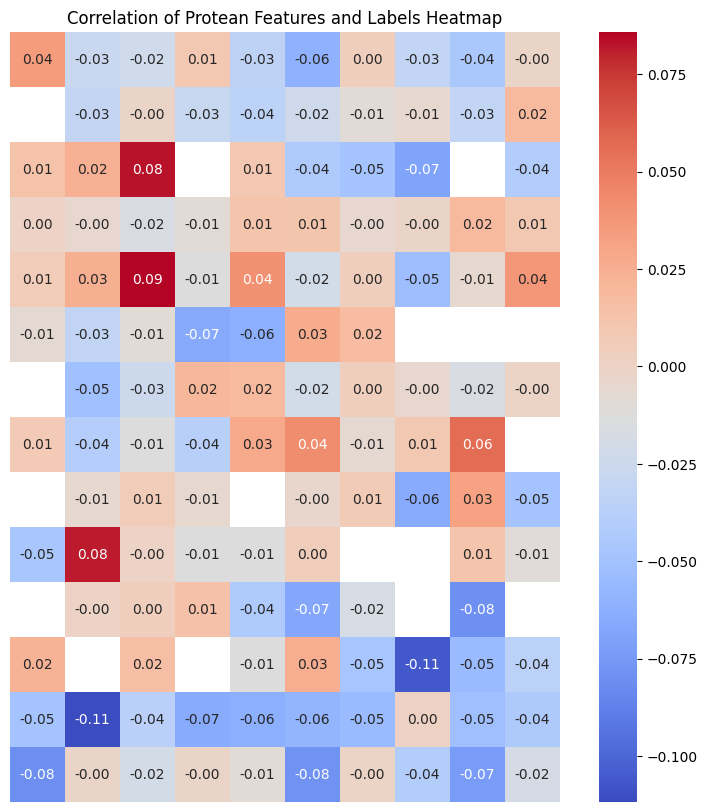}}}
\qquad
\subfloat[IR2VEC Pearson Correlation]{
\label{fig:IR2VECCorr}
\resizebox{.46\textwidth}{!}{\includegraphics{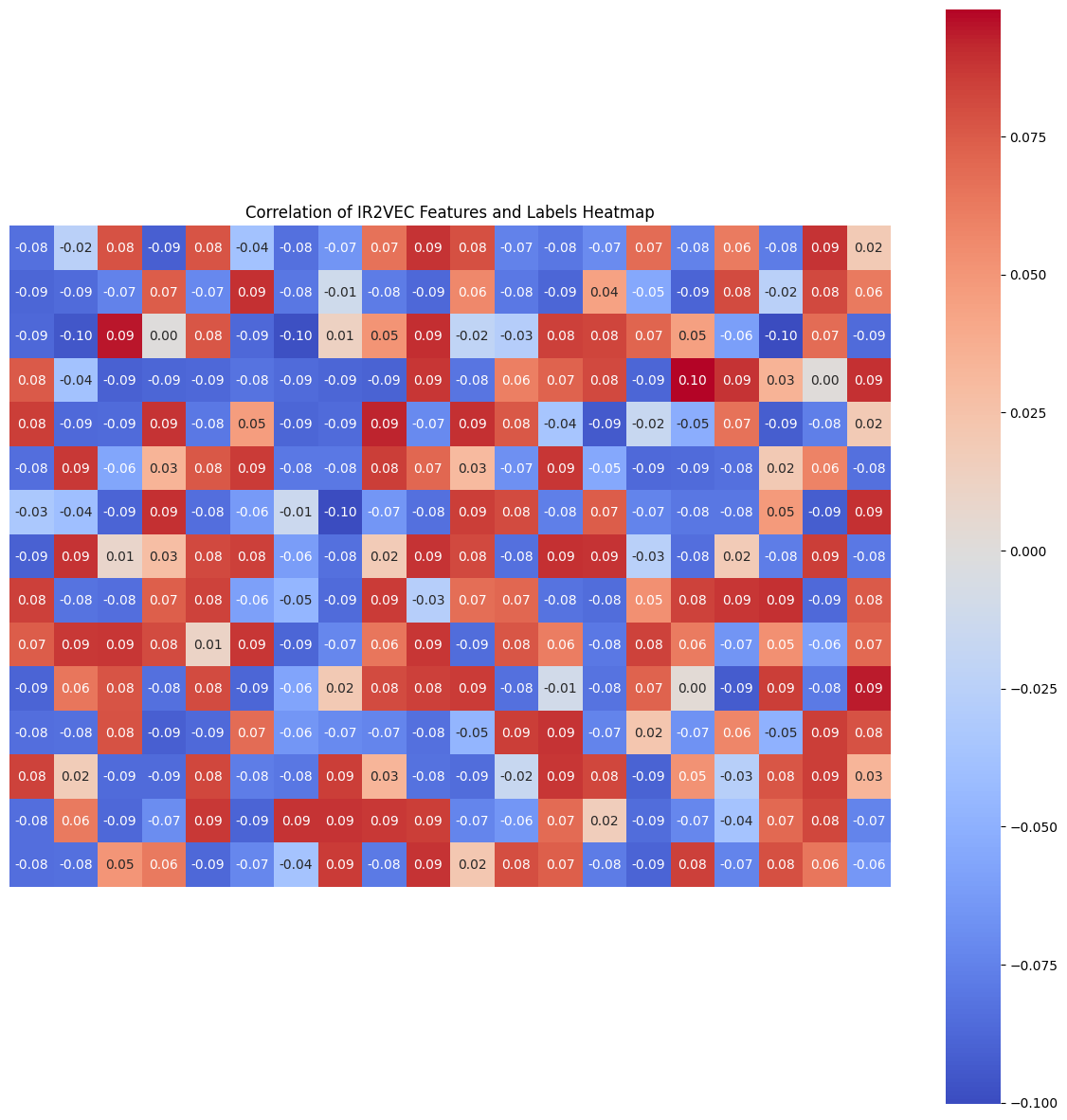}}}
\caption{PFS and IR2VEC Feature Correlation with our Label (Module Speedup against \texttt{-O3})} 
\label{fig:feature_correlation}
\end{figure}

We implement a total number of 141 handcrafted features spanning from module, call graph, function, and loop scopes, and the library of features as a whole or individual scopes can be easily added to the IR2Score model training and inference. Protean feature collection has been implemented as a module pass and registered at the very end of the agile compilation pipeline, and thus, it provides the updated features, as a result of transformed IR at each iteration, to the IR2Score model for inference routines. Table \ref{tab:proteanftlist} depicts a short list of PFS features. \cc{Each scope is further divided into groups where the set of features belonging to a group can be collected together. Each group of features in a scope has a dedicated collector to speed up the collection process by iterating over necessary elements once and dumping registered features under that group at once, if possible. Additionally, Figure \ref{fig:feature_correlation} depicts the correlation of PFS and IR2Vec features, i.e., 141 and 300, against the speedup label we used with IR2Score. Since Pearson has a range of [-1,+1], the more strongly correlated features are either positively or negatively shown with dark blue and red, and although IR2Vec has a higher correlation on average across all of its 300 embeddings, PFS has a dozen strongly correlated features.}

\begin{table*}[!t]
\centering
\small
\begin{tabular}{|>{\centering\arraybackslash}m{1.9cm}| >{\centering\arraybackslash}m{1.2cm}|m{10cm}|}
\hline
\textbf{Scope}                  & \textbf{Features} & \multicolumn{1}{c|}{\textbf{Feature Example}}                                                          \\ \hline
\textbf{Module/global}          & 10                            & FunctionCount,   AverageBBPerFunction, TotalBBCount, TotalInstCount, TotalFunctionCalls, GlobalVariableCount \\ \hline
\textbf{Function}               & 71                            & AverageComponentSize,  SCCSize, IsLocal, CallerHeight,   NumCallsiteInLoop, , AverageStoreInstructionsPerFunction, AverageLoadInstructionsPerFunction, AverageInstructionsPerFunction, , CriticalEdgeCount, TotalEdgeCount, LoopCount, MedianCallsPerFunction, AverageCallsPerFunction, TotalFunctionCalls                     \\ \hline
\textbf{Callee/Caller} & 39                                     & NumCallsiteInLoop,   NestedInlines, SROALosses, IndirectCallPenalty, CallSiteCost, CalleeInstrPerLoop, CallerInstrPerLoop, CalleeAvgNestedLoopLevel, CallerAvgNestedLoopLevel, CalleeAvgVecInstr, CallerAvgVecInstr, CalleeSuccessorPerBlock, CallerSuccessorPerBlock, CalleeInstructionPerBlock, CallerInstructionPerBlock, CalleeMaxCallsiteBlockFreq, CallerMaxCallsiteBlockFreq, CallerEntryBlockFreq, CalleeNumOfCallUsesInLoop, CallerNumOfCallUsesInLoop          \\ \hline
\textbf{Loops}                  & 31                            & LoopSize,   NumLoadInstPerLoopNest, MaxLoopHeight, StepValueInt, TotalEdgeCount, AvgNumLoadInstPerLoop, TotLoopInstCount, NumStoreInstPerLoop, NumLoadInstPerLoop, AvgNumLoadInstPerLoopNest, TotLoopNestInstCount, NumStoreInstPerLoopNest, NumLoadInstPerLoopNest, AvgNumInsts, AvgStoreSetSize, IndVarSetSize, NumPartitions, StepValueInt, FinalIVValueInt, InitialIVValueInt, Size, MaxTripCount, TripCount                \\ \hline
\end{tabular}
\caption{Short list of features in the Protean Feature Set (PFS)}
\label{tab:proteanftlist}
\end{table*}

\section{Experimental Results}
\label{sec:results}

In this section, we evaluate the effectiveness of the Protean compiler on CBench workloads, measuring both performance improvements and the minimal changes in build time. The experimental setup is as follows. 

\paragraph{\taco{Software}} The program runtimes are collected with Linux \texttt{perf}. All the runtime measurements are evaluated single-threaded, and we use the \texttt{numactl} tool to bind the workloads to a unique cpu core and force the memory allocation to come from the same node to make sure there is little noise involved with the measurements. 

\paragraph{\taco{Hardware}} \cc{All experiments are measured on Kunpeng 920 Linux server, with an AArch64 CPU of ARMv8.2 architecture, running at 2.6GHz}. 

\begin{figure*}[!t]
\hspace*{-2em}
\centering
\subfloat[Test]{
\label{fig:test}
\resizebox{.37\textwidth}{!}{\includegraphics{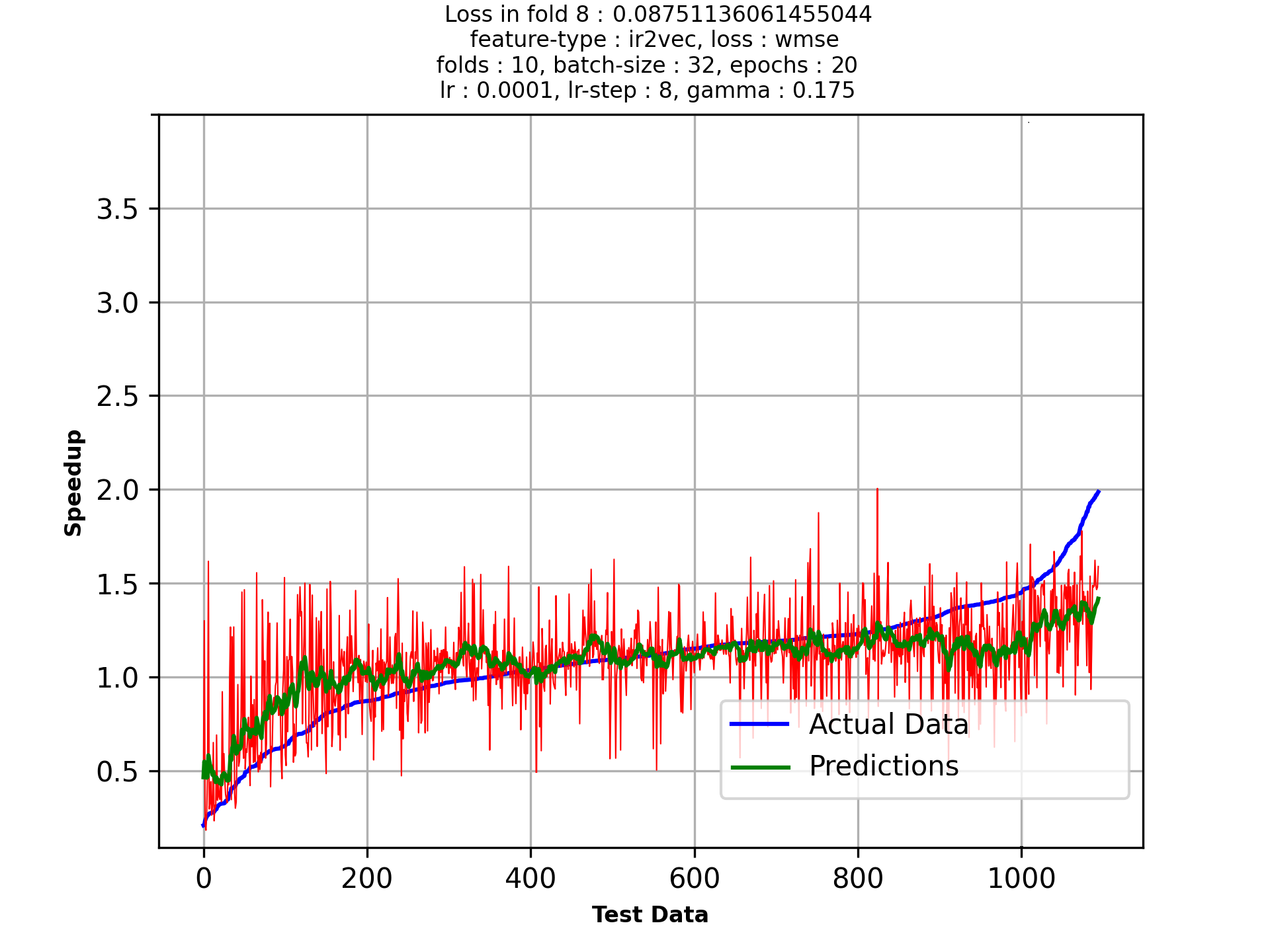}}}
\subfloat[Train]{
\label{fig:training}
\resizebox{.37\textwidth}{!}{\includegraphics{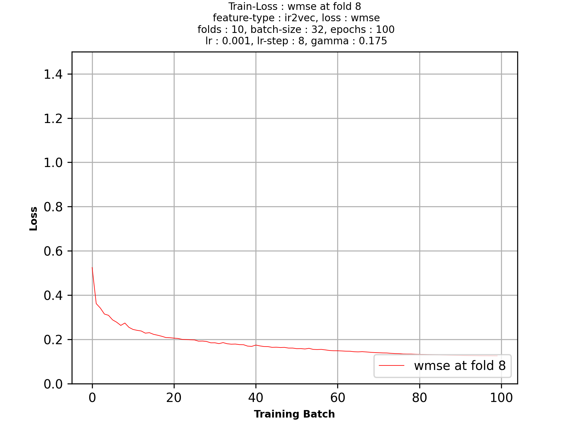}}}
\subfloat[Validation]{
\label{fig:validation}
\resizebox{.37\textwidth}{!}{\includegraphics{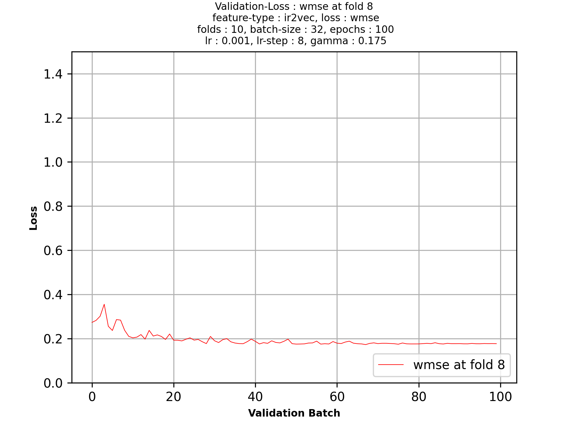}}}
\qquad
\subfloat[Training data recipe length]{
\label{fig:trainingRecipe}
\resizebox{.32\textwidth}{!}{\includegraphics{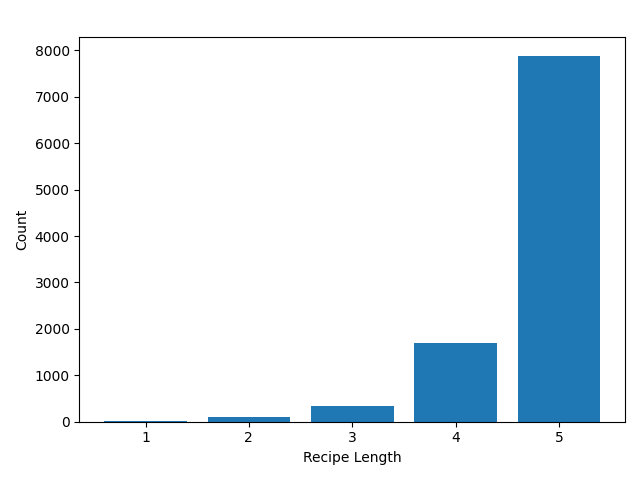}}}
\subfloat[Fraction of available recipe length data vs overall  space]{
\label{fig:trainingRecipeFraction}
\resizebox{.32\textwidth}{!}{\includegraphics{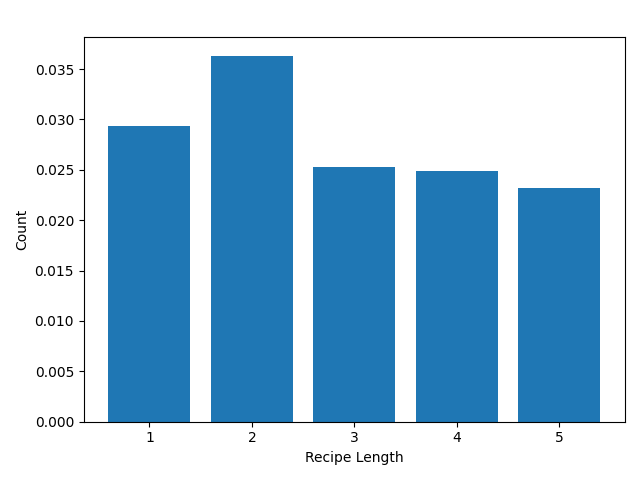}}}
\caption{IR2Score Model Training and Validation} 
\label{fig:ir2score_training_inference}
\end{figure*}

\paragraph{\taco{Methodology}} We run each benchmark five times, and before each measurement run, we flush the system page cache to avoid any perturbation in the collection of our training data. We also repeat the process for all measurements having more than 1\% of variance between the runs to make sure they are all robust and reproducible. For brevity, we omitted the column reporting the variance for each benchmark; however, the mean variance values are 0.58\%, 0.71\%, and 0.49\% for \texttt{-O3}, IR2Score w/ IR2VEC, and IR2Score w/ PFS, respectively.  
To showcase the performance of IR2Score, we chose module-level granularity for our evaluation, i.e., for each benchmark, our agile optimization will spawn a subprocess to iteratively evaluate a recipe for each module, and subsequently, for each resulting IR, we invoke our IR2Score model to evaluate the generated IR. Once each module's subprocesses exit successfully, we iterate over the next module in the program, and the flow continues until all modules are visited and optimized. As mentioned earlier, the Protean compiler, as it is, only focuses on optimizing IR at the {\it opt} stage and leaves the optimizations in the linking stage to the compiler for efficiency. 
\paragraph{IR2Score Training} As discussed in Section \ref{sec:proposed:ir2score}, we chose our Transformer architecture over MLPs or LSTMs as we observed our model has a lower cross-validation loss value of 0.094 vs. 0.958 and 0.125 of the other two, respectively. \cc{We implemented IR2Score with Pytorch v2.6 and the model has 4 layers of Transformer Encoder with 10 attention heads, a max sequence of 5 with mean aggregation of encodings before the regressor and a feedforward network layer dimension of 512. Training was done on a 20-epoch basis with 10 folds of batch sizes of 32 and a learning rate of 0.0001, and a gamma of 0.175, and Adam optimizer. We used a weighted mean-square error (WMSE) as the loss function, and the fold no. 8 was selected as the best overall.  We collected ~12k C/C++ functions from the Polybench and Coral-2 benchmarks, specifically, AMG, lammps, SSSP, BFS, STREAM, STRIDE, and also CoreMark, and cross-validated the data to avoid data leakage}. We have leveraged Linux \texttt{perf} to identify runtimes of functions \cc{and the number of iterations/recursions}, and thus, the summation of all functions in a module constructs the total runtime of a module in a program. Note that the one tricky scenario where a function is inlined can be revealed by inspecting the call graph of the executable and updating the list of functions that are contributing to the current module's runtime post-transformation of a recipe. Similarly, the recipe (subsequence) speedup can be determined by means of a division of the LLVM's \texttt{-O3} and the module runtime.  \taco{Figure \ref{fig:ir2score_training_inference} depicts the representation of the IR2Score model and the training/validation graphs. We noticed that the model struggled to learn the ground truth speedup curve for samples with speedup less than 0.8 and more than 1.5. Therefore, we used a weighted mean squared loss function to train the model with higher weights assigned to the samples in those regions, which has helped significantly in increasing the accuracy of the model as shown in Figure \ref{fig:test}. The test was excluded from the training set and thus, it shows a good generalization across by closely following the actual values. Additionally, Figures \ref{fig:trainingRecipe} and \ref{fig:trainingRecipeFraction} reveal our current IR2Score training data filtered by the recipe length, i.e., 1 to 5, and the fraction of generated/available training data versus full exhaustive phase ordering search of the training applications under analysis, respectively.}

\begin{figure*}[!t]
\hspace*{-1em}
\centering
\subfloat[SA+GA Recommender Benefits]{
\label{fig:genetic}
\resizebox{.47\textwidth}{.37\textwidth}{\includegraphics{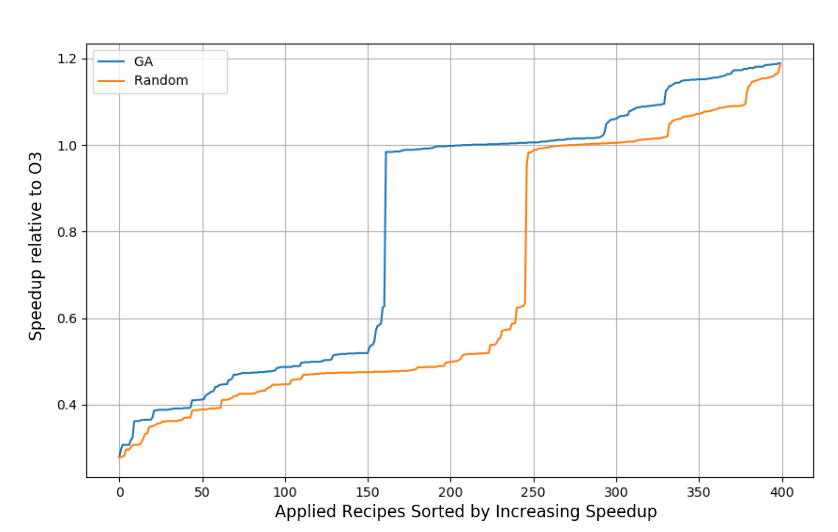}}}
\subfloat[\centering{CBench Performance Gains} \newline (Dashed Green: Mean, Solid Orange: Median)]{
\label{fig:explore-exploit}
\resizebox{.52\textwidth}{!}{\includegraphics{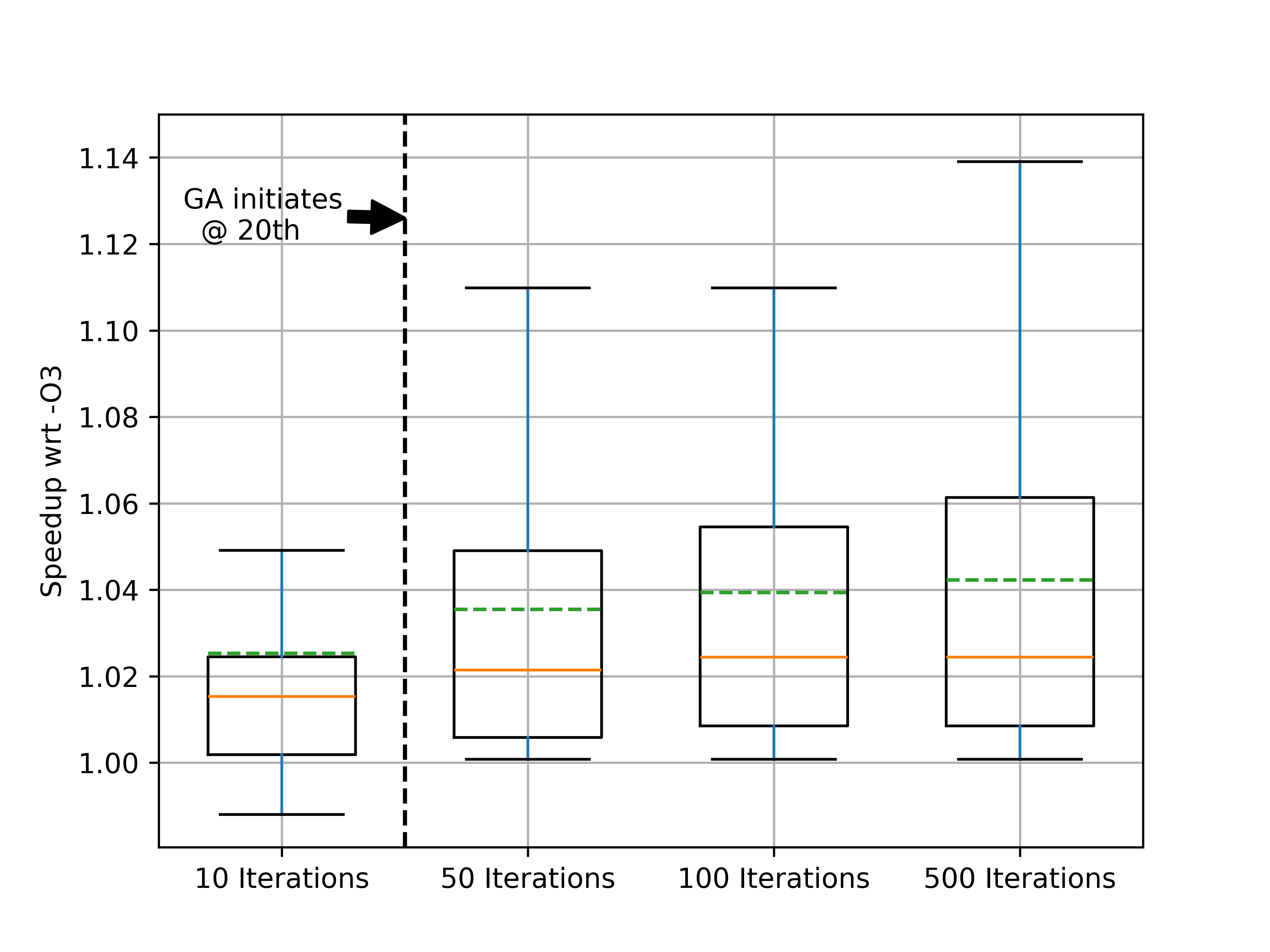}}}
\caption{Protean Compiler Exploration vs. Exploitation Analysis} 
\label{fig:exploration-exploitation}
\end{figure*}

\paragraph{IR2Score Deployment} We generate Ahead-Of-Time (AOT) models of IR2Score models when compiled into C++ libraries and leverage MLGO's \cite{trofin2021mlgo} \texttt{ModelRunner} classes and ACPO's \cite{ashouri2023acpo} framework to deploy our models in LLVM 19. Once the model's AOT is linked with LLVM, inference routines become function calls; therefore, the main overhead becomes only the feature collection process. The next section will provide build-time analysis for our two IR2SCore models. Additionally, we included an \textit{early exit condition} during the inference routine, which checks if the IR remains unchanged for a certain number of times as a result of converging to a (locally) optimal instance before the set number of iterations is met. This also speeds up the cases where the modules are too small to realize meaningful changes as a result of changing recipes. \taco{This parameter is set to 100 iterations by default}. There are a number of hyperparameters used with running Protean, \taco{including a [0-100] range for the temperature and a geometric cooling algorithm to traverse from the maximum temperature to the minimum, and using 20, 0.05, and 0.95 as initial population size, mutation, and crossover rate for our genetic algorithm (GA), respectively. Figure \ref{fig:genetic} showcases the benefits of our GA recommender system when paired with SA engine in speeding up the search. In this result, we have tested both approaches on one selected CBench from each category (similar to identifying the recipe length in the previous section) and sorted the speedup values found by exploring the space. The Y axis represents the speedup relative to \texttt{-O3}, and 0.3 and 1.2 are the lowest and highest available amounts of speedup, respectively. We can experimentally deduce that the addition of GA to our SA clearly moves the line earlier and to the left of the figure.}  

In the following subsections, we discuss the impact of the Protean compiler on workload performance as we vary the number of iterations. We also discuss the build time as a function of iteration count to discuss trade-offs to be considered when using the Protean compiler.

\subsection{Performance Analysis of Protean Compiler}
\label{sec:results:performance}

\subsubsection{CBench}
\label{sec:results:performance:cbench}

Table \ref{tab:cbench_perf} showcases CBench performance results with IR being evaluated by our IR2Score model. For evaluation purposes, we examined the impact of reducing the rate of temperature decrease, which effectively increases the number of iterations before the final result stabilizes. We controlled the temperature gradient to explore 10, 50, 100, and 500 iterations. In each case, we evaluated the performance of the Protean compiler using the IR2Score model that used different feature sets: IR2VEC~\cite{venkatakeerthy2020ir2vec} and Protean Feature Set (PFS). We observe that the Protean compiler can speed up CBench application on average from 2.2\% up to 4.1\%, with certain applications reaching double-digit speedup values of up to 15\%. Additionally, PFS feature collection consistently outperforms IR2VEC, and we believe one contributing factor is that the authors of IR2VEC had trained the embeddings on SPEC CPU applications, so there is a possibility of underfitting here. \taco{Figure \ref{fig:explore-exploit} showcases the boxplot distributions of our speedup values. Each boxplot depicts the mean (in dashed green line) and median (in solid orange line) speed-up values found under each category when Protean was used with PFS on CBench. We observe that our genetic algorithm (currently initiated at the 20th iteration after observing its initial population's performance) is able to gain the biggest increase in performance between 10 and 50 iterations, where we jump from 2.2\% to 3.3\% and 2.5\% to 3.5\% for models trained with IR2VEC and PFS, respectively. This reveals the benefits of leveraging GS together with SA in the exploration policy of Protean, as shown earlier in Figure \ref{fig:genetic} from another angle. }

\begin{table*}[!t]
\centering
\hspace*{-1em}
\footnotesize
\begin{tabular}{|lc|cc?cc|cc|cc|}
\multicolumn{4}{r?}{\textbf{GA initiates @ 20th iteration} $\longrightarrow$}                                                        & \multicolumn{6}{c}{\textbf{}}           \\ 
\hline
\multicolumn{1}{|c|}{\multirow{2}{*}{\textbf{Benchmark}}} & \multirow{2}{*}{\textbf{-O3 (s)}} & \multicolumn{2}{c?}{\textbf{Speedup (10 iters)}}                      & \multicolumn{2}{c|}{\textbf{50 Iterations}}                      & \multicolumn{2}{c|}{\textbf{100 Iterations}}                    & \multicolumn{2}{c|}{\textbf{500 Iterations}}                     \\ \cline{3-10} 
\multicolumn{1}{|c|}{}                                  &                                     & \multicolumn{1}{c|}{\textbf{IR2VEC}} & \textbf{PFS} & \multicolumn{1}{c|}{\textbf{IR2VEC}} & \textbf{PFS} & \multicolumn{1}{c|}{\textbf{IR2VEC}} & \textbf{PFS} & \multicolumn{1}{c|}{\textbf{IR2VEC}} & \textbf{PFS} \\ \hline
\multicolumn{1}{|l|}{\textbf{automotive\_bitcount}}     & 1.655                              & \multicolumn{1}{c|}{1.005}              & 1.023                 & \multicolumn{1}{c|}{1.011}              & 1.023                 & \multicolumn{1}{c|}{1.022}              & 1.028                 & \multicolumn{1}{c|}{1.028}              & 1.028                 \\ \hline
\multicolumn{1}{|l|}{\textbf{automotive\_susan\_c}}     & 4.689                               & \multicolumn{1}{c|}{1.015}              & 1.036                 & \multicolumn{1}{c|}{1.024}              & 1.037                 & \multicolumn{1}{c|}{1.026}              & 1.037                 & \multicolumn{1}{c|}{1.033}              & 1.037                 \\ \hline
\multicolumn{1}{|l|}{\textbf{automotive\_susan\_e}}     & 4.090                              & \multicolumn{1}{c|}{1.041}              & 1.022                 & \multicolumn{1}{c|}{1.080}              & 1.045                 & \multicolumn{1}{c|}{1.080}              & 1.094                 & \multicolumn{1}{c|}{1.080}              & 1.094                 \\ \hline
\multicolumn{1}{|l|}{\textbf{automotive\_susan\_s}}     & 3.666                             & \multicolumn{1}{c|}{0.995}              & 1.009                 & \multicolumn{1}{c|}{0.995}              & 1.009                 & \multicolumn{1}{c|}{0.995}              & 1.009                 & \multicolumn{1}{c|}{0.995}              & 1.009                 \\ \hline
\multicolumn{1}{|l|}{\textbf{bzip2d}}                   & 5.137                              & \multicolumn{1}{c|}{0.998}              & 1.049                 & \multicolumn{1}{c|}{1.000}              & 1.049                 & \multicolumn{1}{c|}{1.000}              & 1.049                 & \multicolumn{1}{c|}{1.000}              & 1.049                 \\ \hline
\multicolumn{1}{|l|}{\textbf{bzip2e}}                   & 5.028                              & \multicolumn{1}{c|}{0.997}              & 1.000                 & \multicolumn{1}{c|}{0.997}              & 1.002                 & \multicolumn{1}{c|}{0.997}              & 1.002                 & \multicolumn{1}{c|}{0.997}              & 1.002                 \\ \hline
\multicolumn{1}{|l|}{\textbf{consumer\_jpeg\_c}}        & 5.845                              & \multicolumn{1}{c|}{1.006}              & 0.988                 & \multicolumn{1}{c|}{1.042}              & 1.005                 & \multicolumn{1}{c|}{1.042}              & 1.005                 & \multicolumn{1}{c|}{1.042}              & 1.005                 \\ \hline
\multicolumn{1}{|l|}{\textbf{consumer\_jpeg\_d}}        & 15.755                              & \multicolumn{1}{c|}{1.017}              & 1.024                 & \multicolumn{1}{c|}{1.044}              & 1.072                 & \multicolumn{1}{c|}{1.058}              & 1.078                 & \multicolumn{1}{c|}{1.073}              & 1.078                 \\ \hline
\multicolumn{1}{|l|}{\textbf{consumer\_lame}}           & 5.584                              & \multicolumn{1}{c|}{1.008}              & 1.017                 & \multicolumn{1}{c|}{1.008}              & 1.017                 & \multicolumn{1}{c|}{1.008}              & 1.017                 & \multicolumn{1}{c|}{1.008}              & 1.017                 \\ \hline
\multicolumn{1}{|l|}{\textbf{consumer\_mad}}            & 6.972                               & \multicolumn{1}{c|}{1.027}              & 1.020                 & \multicolumn{1}{c|}{1.027}              & 1.020                 & \multicolumn{1}{c|}{1.027}              & 1.020                 & \multicolumn{1}{c|}{1.027}              & 1.020                 \\ \hline
\multicolumn{1}{|l|}{\textbf{consumer\_tiffdither}}     & 7.160                              & \multicolumn{1}{c|}{0.990}              & 1.003                 & \multicolumn{1}{c|}{0.990}              & 1.003                 & \multicolumn{1}{c|}{0.990}              & 1.003                 & \multicolumn{1}{c|}{0.990}              & 1.003                 \\ \hline
\multicolumn{1}{|l|}{\textbf{consumer\_tiffmedian}}     & 5.587                              & \multicolumn{1}{c|}{1.031}              & 1.013                 & \multicolumn{1}{c|}{1.031}              & 1.013                 & \multicolumn{1}{c|}{1.031}              & 1.013                 & \multicolumn{1}{c|}{1.031}              & 1.013                 \\ \hline
\multicolumn{1}{|l|}{\textbf{network\_dijkstra}}        & 0.639                             & \multicolumn{1}{c|}{1.081}              & 1.075                 & \multicolumn{1}{c|}{1.081}              & 1.075                 & \multicolumn{1}{c|}{1.081}              & 1.075                 & \multicolumn{1}{c|}{1.100}              & 1.139                 \\ \hline
\multicolumn{1}{|l|}{\textbf{network\_patricia}}        & 2.796                              & \multicolumn{1}{c|}{1.007}              & 1.008                 & \multicolumn{1}{c|}{1.007}              & 1.008                 & \multicolumn{1}{c|}{1.007}              & 1.008                 & \multicolumn{1}{c|}{1.007}              & 1.008                 \\ \hline
\multicolumn{1}{|l|}{\textbf{office\_ghostscript}}      & 1.015                              & \multicolumn{1}{c|}{1.000}              & 1.003                 & \multicolumn{1}{c|}{1.000}              & 1.013                 & \multicolumn{1}{c|}{1.000}              & 1.013                 & \multicolumn{1}{c|}{1.000}              & 1.013                 \\ \hline
\multicolumn{1}{|l|}{\textbf{office\_rsynth}}           & 1.110                             & \multicolumn{1}{c|}{1.110}              & 1.110                 & \multicolumn{1}{c|}{1.110}              & 1.110                 & \multicolumn{1}{c|}{1.110}              & 1.110                 & \multicolumn{1}{c|}{1.110}              & 1.110                 \\ \hline
\multicolumn{1}{|l|}{\textbf{office\_stringsearch1}}    & 4.223                              & \multicolumn{1}{c|}{0.999}              & 1.002                 & \multicolumn{1}{c|}{0.999}              & 1.004                 & \multicolumn{1}{c|}{0.999}              & 1.004                 & \multicolumn{1}{c|}{1.001}              & 1.004                 \\ \hline
\multicolumn{1}{|l|}{\textbf{security\_blowfish\_d}}    & 8.497                              & \multicolumn{1}{c|}{1.017}              & 1.004                 & \multicolumn{1}{c|}{1.017}              & 1.004                 & \multicolumn{1}{c|}{1.022}              & 1.009                 & \multicolumn{1}{c|}{1.022}              & 1.009                 \\ \hline
\multicolumn{1}{|l|}{\textbf{security\_blowfish\_e}}    & 8.516                               & \multicolumn{1}{c|}{1.025}              & 0.996                 & \multicolumn{1}{c|}{1.025}              & 1.036                 & \multicolumn{1}{c|}{1.025}              & 1.036                 & \multicolumn{1}{c|}{1.025}              & 1.036                 \\ \hline
\multicolumn{1}{|l|}{\textbf{security\_pgp\_d}}         & 1.111                             & \multicolumn{1}{c|}{1.101}              & 1.111                 & \multicolumn{1}{c|}{1.158}              & 1.134                 & \multicolumn{1}{c|}{1.158}              & 1.157                 & \multicolumn{1}{c|}{1.158}              & 1.157                 \\ \hline
\multicolumn{1}{|l|}{\textbf{security\_pgp\_e}}         & 1.107                             & \multicolumn{1}{c|}{1.099}              & 1.107                 & \multicolumn{1}{c|}{1.099}              & 1.107                 & \multicolumn{1}{c|}{1.099}              & 1.107                 & \multicolumn{1}{c|}{1.099}              & 1.107                 \\ \hline
\multicolumn{1}{|l|}{\textbf{security\_sha}}            & 6.215                               & \multicolumn{1}{c|}{1.004}              & 1.022                 & \multicolumn{1}{c|}{1.055}              & 1.049                 & \multicolumn{1}{c|}{1.055}              & 1.056                 & \multicolumn{1}{c|}{1.061}              & 1.065                 \\ \hline
\multicolumn{1}{|l|}{\textbf{telecom\_adpcm\_c}}        & 5.204                              & \multicolumn{1}{c|}{1.000}              & 0.995                 & \multicolumn{1}{c|}{1.000}              & 1.001                 & \multicolumn{1}{c|}{1.000}              & 1.001                 & \multicolumn{1}{c|}{1.000}              & 1.001                 \\ \hline
\multicolumn{1}{|l|}{\textbf{telecom\_adpcm\_d}}        & 4.223                             & \multicolumn{1}{c|}{0.991}              & 1.001                 & \multicolumn{1}{c|}{1.039}              & 1.050                 & \multicolumn{1}{c|}{1.050}              & 1.050                 & \multicolumn{1}{c|}{1.050}              & 1.050                 \\ \hline
\multicolumn{1}{|l|}{\textbf{telecom\_CRC32}}           & 2.769                             & \multicolumn{1}{c|}{1.001}              & 0.997                 & \multicolumn{1}{c|}{1.001}              & 1.001                 & \multicolumn{1}{c|}{1.001}              & 1.001                 & \multicolumn{1}{c|}{1.001}              & 1.001                 \\ \hline
\multicolumn{2}{|c|}{\textbf{Geomean}}                                                        & \multicolumn{1}{c|}{\textbf{1.022}}     & \textbf{1.025}        & \multicolumn{1}{c|}{\textbf{1.033}}     & \textbf{1.035}        & \multicolumn{1}{c|}{\textbf{1.035}}     & \textbf{1.039}        & \multicolumn{1}{c|}{\textbf{1.037}}     & \textbf{1.041}        \\ \hline
\end{tabular}
\caption{CBench Performance }
\label{tab:cbench_perf}
\end{table*}

\subsubsection{SPEC CPU 2017}
\label{sec:results:performance:spec}

\taco{Table \ref{tab:spec_perf} presents SPEC CPU 2017 performance results (Integer suite). We run Protean compiler with IR2VEC for 30 iterations (GA initiates at the 20th iteration) and we capture the binary size and the execution times at two scenarios -- single-copy and 64-copy runs. This setting can be supplied to \texttt{runcpu} by means of \texttt{$--$copies=\{1,64}\} to launch multiple instances of execution runs to measure throughput by running identical, concurrent jobs. In this scenario, the slowest job becomes the benchmark run, and we run each benchmark three times and use a trimmed mean to get the execution time values. We focus on its 9 C/C++ applications, and Table \ref{tab:spec_perf} shows the results. As shown, Protean compiler is able to speed up SPEC benchmarks (geometric mean value) by 1\% and 1.1\% on 1-copy and 64-copy run scenarios, respectively. This performance gain comes at the code size increase of ~1.5\%, although a number of benchmarks, i.e., 531.deepsjeng and 541.leela benefited from a reduced binary code size as well. One interesting observation is for 557.xz, a data compression benchmark, which benefits more with a higher number of parallel execution runs, i.e., 1.3\% vs 6.5\%. Protean's phase ordering, despite increasing the binary, has benefited from inlining of a few hot functions, and thus, a better utilization of repetitive function calls led to higher throughput and improved instruction caching when 64 copies (jobs) were simultaneously run on our server.}

\begin{table*}[!t]
\centering
\hspace*{-1em}
\footnotesize
\begin{tabular}{|l|ccc|ccc|ccc|}
\hline
\multicolumn{1}{|c|}{\multirow{2}{*}{\textbf{Benchmark}}} & \multicolumn{3}{c|}{\textbf{-O3}}                                                                                                                  & \multicolumn{3}{c|}{\textbf{Protean (w/IR2VEC) -- 30 Iters}}                                                                                                 & \multicolumn{3}{c|}{\textbf{Protean Speedup}}                                                                                                 \\ \cline{2-10} 
\multicolumn{1}{|c|}{}                                    & \multicolumn{1}{A{.8cm}|}{\textbf{Binary Size (KB)}} & \multicolumn{1}{A{1cm}|}{\textbf{Runtime 1 Copy (s)}} & \multicolumn{1}{A{1cm}|}{\textbf{Runtime 64 Copy (s)}} & \multicolumn{1}{A{.8cm}|}{\textbf{Binary Size (KB)}} & \multicolumn{1}{A{1cm}|}{\textbf{Runtime 1 Copy (s)}} & \multicolumn{1}{A{1cm}|}{\textbf{Runtime 64 Copy (s)}} & \multicolumn{1}{A{.9cm}|}{\textbf{Size Bloat (\%)}} & \multicolumn{1}{A{1cm}|}{\textbf{Runtime 1 Copy (\%)}} & \multicolumn{1}{A{1cm}|}{\textbf{Runtime 64 Copy (\%)}} \\ \hline
\textbf{500.perlbench}                                 & \multicolumn{1}{c|}{4748}                      & \multicolumn{1}{c|}{503.31}                    & \multicolumn{1}{c|}{591.16}                     & \multicolumn{1}{c|}{4901}                      & \multicolumn{1}{c|}{501.48}                    & \multicolumn{1}{c|}{591.16}                     & \multicolumn{1}{c|}{1.032}                      & \multicolumn{1}{c|}{1.004}                     & \multicolumn{1}{c|}{1.000}                      \\ \hline
\textbf{502.gcc}                                       & \multicolumn{1}{c|}{18679}                     & \multicolumn{1}{c|}{348.86}                    & \multicolumn{1}{c|}{705.65}                     & \multicolumn{1}{c|}{20289}                     & \multicolumn{1}{c|}{349.20}                    & \multicolumn{1}{c|}{706.36}                     & \multicolumn{1}{c|}{1.086}                      & \multicolumn{1}{c|}{0.999}                     & \multicolumn{1}{c|}{0.999}                      \\ \hline
\textbf{505.mcf}                                       & \multicolumn{1}{c|}{33}                        & \multicolumn{1}{c|}{439.59}                    & \multicolumn{1}{c|}{1171.81}                    & \multicolumn{1}{c|}{33}                        & \multicolumn{1}{c|}{439.01}                    & \multicolumn{1}{c|}{1166.30}                    & \multicolumn{1}{c|}{1.000}                      & \multicolumn{1}{c|}{1.001}                     & \multicolumn{1}{c|}{1.005}                      \\ \hline
\textbf{520.omnetpp}                                   & \multicolumn{1}{c|}{2524}                      & \multicolumn{1}{c|}{347.18}                    & \multicolumn{1}{c|}{346.16}                     & \multicolumn{1}{c|}{2524}                      & \multicolumn{1}{c|}{346.16}                    & \multicolumn{1}{c|}{346.16}                     & \multicolumn{1}{c|}{1.000}                      & \multicolumn{1}{c|}{1.003}                     & \multicolumn{1}{c|}{1.000}                      \\ \hline
\textbf{523.xalancbmk}                                 & \multicolumn{1}{c|}{5565}                      & \multicolumn{1}{c|}{164.99}                    & \multicolumn{1}{c|}{362.76}                     & \multicolumn{1}{c|}{5697}                      & \multicolumn{1}{c|}{157.69}                    & \multicolumn{1}{c|}{354.26}                     & \multicolumn{1}{c|}{1.024}                      & \multicolumn{1}{c|}{1.046}                     & \multicolumn{1}{c|}{1.024}                      \\ \hline
\textbf{525.x264}\tablefootnote{\taco{ x265 comprises of 3 binaries: imagevalidate, ldecod, and x264. The binary size reflects the summation of the three.} 
}                                  & \multicolumn{1}{c|}{1374}                      & \multicolumn{1}{c|}{226.77}                    & \multicolumn{1}{c|}{236.48}                     & \multicolumn{1}{c|}{1371}                      & \multicolumn{1}{c|}{227.77}                    & \multicolumn{1}{c|}{236.48}                     & \multicolumn{1}{c|}{0.998}                      & \multicolumn{1}{c|}{0.996}                     & \multicolumn{1}{c|}{1.000}                      \\ \hline
\textbf{531.deepsjeng}                                 & \multicolumn{1}{c|}{121}                       & \multicolumn{1}{c|}{280.36}                    & \multicolumn{1}{c|}{287.17}                     & \multicolumn{1}{c|}{117}                       & \multicolumn{1}{c|}{278.00}                    & \multicolumn{1}{c|}{287.17}                     & \multicolumn{1}{c|}{0.964}                      & \multicolumn{1}{c|}{1.008}                     & \multicolumn{1}{c|}{1.000}                      \\ \hline
\textbf{541.leela}                                     & \multicolumn{1}{c|}{140}                       & \multicolumn{1}{c|}{481.81}                    & \multicolumn{1}{c|}{488.21}                     & \multicolumn{1}{c|}{138}                       & \multicolumn{1}{c|}{471.99}                    & \multicolumn{1}{c|}{482.70}                     & \multicolumn{1}{c|}{0.986}                      & \multicolumn{1}{c|}{1.021}                     & \multicolumn{1}{c|}{1.011}                      \\ \hline
\textbf{557.xz}                                        & \multicolumn{1}{c|}{179}                       & \multicolumn{1}{c|}{426.50}                    & \multicolumn{1}{c|}{744.57}                     & \multicolumn{1}{c|}{187}                       & \multicolumn{1}{c|}{421.09}                    & \multicolumn{1}{c|}{699.24}                     & \multicolumn{1}{c|}{1.043}                      & \multicolumn{1}{c|}{1.013}                     & \multicolumn{1}{c|}{1.065}                      \\ \hline
\textbf{Average}                                          & 3707                                           & 358                                            & 548                                             & 3917                                           & 355                                            & 541                                             & \textbf{1.014}                                  & \textbf{1.010}                                 & \textbf{1.011}                                  \\ \cline{1-10}
\end{tabular}
\caption{SPEC CPU 2017 Performance (30 Iterations w/ IR2VEC, GA initiates @ 20th iteration)  }
\label{tab:spec_perf}
\end{table*}

\subsection{Protean Compilation Overhead Analysis}
\label{sec:results:buildtime}

\subsubsection{CBench}
\label{sec:results:buildtime:cbench}

Table \ref{tab:cbench_build} showcases the single-core build-time overhead of our IR2Score models when deployed on CBench. The values reported under each column represent the added overhead in seconds to the baseline build, i.e., LLVM's \texttt{-O3}. For instance, we can see that building \texttt{automotive\_bitcount} benchmark has an added overhead of 2 to only 36 seconds when the IR2Score model is deployed from 10 to 500 iterations using either IR2VEC or Protean features. As shown, the overhead of running the end-to-end Protean compiler is minimal when it comes to phase-ordering of optimizations. There are two main reasons for that: (1) IR2Score's early stop condition exits the agile optimization of a module when it has already converged for a certain number of iterations, and (2) Protean's agile driver applies only IR passes using its ML-based prediction model and exploration strategy, leaving heavy load linker passes intact, and thus increasing iterations doesn't exponentially increase the total build time. 

One interesting observation for the 500-iteration case in Table \ref{tab:cbench_build} is that the PFS is clearly lower than IR2VEC and shows its superiority over IR2VEC, as IR2Score trained with the former converges earlier on better solutions more rapidly. \taco{This is the direct result of using our early exit hyperparameter, and based on its current setting (at 100), it can occur at iterations between 100 and 500. It reveals that our engine has converged on a local/global optimum and has not found a better solution for the past 100 iterations.} For instance, average build time overhead doesn't increase much from 100 to 500 iterations for PFS, as on average, CBench test cases are converging earlier, and around 145 iterations, whereas for IR2VEC's case, the needed average iteration was recorded at around 310 iterations. \taco{The final iteration at which the Protean compiler stopped compilation of each benchmark is reported in Table \ref{tab:cbench_build_parallel}.}

\begin{table*}[!t]
\centering
\footnotesize
\begin{tabular}{|l|>{\centering\arraybackslash}m{.6cm}|cc?cc?cc?cc|}
\hline
\multicolumn{1}{|c|}{\multirow{2}{*}{\textbf{Benchmark}}} & \multirow{2}{*}{\textbf{-O3}} & \multicolumn{2}{c|}{\textbf{10 Iterations}}                 & \multicolumn{2}{c|}{\textbf{50 Iterations}}                 & \multicolumn{2}{c|}{\textbf{100 Iterations}}               & \multicolumn{2}{c|}{\textbf{500 Iterations}}                \\ \cline{3-10} 
\multicolumn{1}{|c|}{}                                    &                                         & \multicolumn{1}{c|}{\textbf{IR2VEC}} & \textbf{PFS} & \multicolumn{1}{c|}{\textbf{IR2VEC}} & \textbf{PFS} & \multicolumn{1}{c|}{\textbf{IR2VEC}} & \textbf{PFS} & \multicolumn{1}{c|}{\textbf{IR2VEC}} & \textbf{PFS} \\ \hline
\textbf{automotive\_bitcount}                             & 1.50                              & \multicolumn{1}{c|}{2.53}            & 2.47                & \multicolumn{1}{c|}{7.49}            & 7.37                & \multicolumn{1}{c|}{13.47}           & 13.73               & \multicolumn{1}{c|}{36.75}           & 17.92               \\ \hline
\textbf{automotive\_susan\_c}                             & 5.91                              & \multicolumn{1}{c|}{14.36}           & 13.76               & \multicolumn{1}{c|}{48.66}           & 47.67               & \multicolumn{1}{c|}{76.80}           & 100.82              & \multicolumn{1}{c|}{250.02}          & 139.31              \\ \hline
\textbf{automotive\_susan\_e}                             & 5.12                              & \multicolumn{1}{c|}{13.94}           & 15.94               & \multicolumn{1}{c|}{48.55}           & 47.40               & \multicolumn{1}{c|}{76.46}           & 100.72              & \multicolumn{1}{c|}{249.74}          & 160.62              \\ \hline
\textbf{automotive\_susan\_s}                             & 5.01                             & \multicolumn{1}{c|}{14.13}           & 17.13               & \multicolumn{1}{c|}{48.72}           & 56.73               & \multicolumn{1}{c|}{91.52}           & 117.12              & \multicolumn{1}{c|}{249.97}          & 164.14              \\ \hline
\textbf{bzip2d}                                           & 7.13                             & \multicolumn{1}{c|}{58.83}           & 66.43               & \multicolumn{1}{c|}{220.73}          & 231.44              & \multicolumn{1}{c|}{380.70}          & 485.49              & \multicolumn{1}{c|}{1398.27}         & 571.86              \\ \hline
\textbf{bzip2e}                                           & 6.99                              & \multicolumn{1}{c|}{58.61}           & 58.25               & \multicolumn{1}{c|}{221.13}          & 271.21              & \multicolumn{1}{c|}{404.12}          & 439.60              & \multicolumn{1}{c|}{1480.60}         & 571.45              \\ \hline
\textbf{consumer\_jpeg\_c}                                & 12.55                             & \multicolumn{1}{c|}{71.36}           & 71.32               & \multicolumn{1}{c|}{276.67}          & 258.96              & \multicolumn{1}{c|}{465.01}          & 542.63              & \multicolumn{1}{c|}{1674.40}         & 697.23              \\ \hline
\textbf{consumer\_jpeg\_d}                                & 12.52                             & \multicolumn{1}{c|}{72.76}           & 74.29               & \multicolumn{1}{c|}{269.90}          & 257.81              & \multicolumn{1}{c|}{467.72}          & 532.25              & \multicolumn{1}{c|}{1626.28}         & 708.17              \\ \hline
\textbf{consumer\_lame}                                   & 11.70                             & \multicolumn{1}{c|}{56.06}           & 55.93               & \multicolumn{1}{c|}{201.26}          & 204.69              & \multicolumn{1}{c|}{360.62}          & 406.06              & \multicolumn{1}{c|}{1260.80}         & 510.25              \\ \hline
\textbf{consumer\_mad}                                    & 9.86                             & \multicolumn{1}{c|}{45.77}           & 47.77               & \multicolumn{1}{c|}{182.94}          & 178.89              & \multicolumn{1}{c|}{307.71}          & 317.73              & \multicolumn{1}{c|}{1081.87}         & 425.83              \\ \hline
\textbf{consumer\_tiffdither}                             & 10.23                             & \multicolumn{1}{c|}{74.72}           & 74.62               & \multicolumn{1}{c|}{311.63}          & 288.95              & \multicolumn{1}{c|}{530.29}          & 538.54              & \multicolumn{1}{c|}{1895.54}         & 714.60              \\ \hline
\textbf{consumer\_tiffmedian}                             & 10.68                             & \multicolumn{1}{c|}{77.38}           & 77.57               & \multicolumn{1}{c|}{312.55}          & 305.14              & \multicolumn{1}{c|}{549.20}          & 558.14              & \multicolumn{1}{c|}{1909.24}         & 737.66              \\ \hline
\textbf{network\_dijkstra}                                & 0.84                              & \multicolumn{1}{c|}{0.99}            & 0.91                & \multicolumn{1}{c|}{2.31}            & 2.60                & \multicolumn{1}{c|}{3.75}            & 4.04                & \multicolumn{1}{c|}{12.58}           & 5.45                \\ \hline
\textbf{network\_patricia}                                & 0.78                              & \multicolumn{1}{c|}{1.48}            & 1.54                & \multicolumn{1}{c|}{3.75}            & 3.94                & \multicolumn{1}{c|}{6.99}            & 7.28                & \multicolumn{1}{c|}{25.33}           & 9.64                \\ \hline
\textbf{office\_ghostscript}                              & 59.50                             & \multicolumn{1}{c|}{103.20}          & 103.30              & \multicolumn{1}{c|}{735.86}          & 665.77              & \multicolumn{1}{c|}{1062.70}         & 1121.52             & \multicolumn{1}{c|}{3310.81}         & 1470.54             \\ \hline
\textbf{office\_rsynth}                                   & 2.65                              & \multicolumn{1}{c|}{6.62}            & 6.44                & \multicolumn{1}{c|}{24.01}           & 23.77               & \multicolumn{1}{c|}{39.85}           & 41.54               & \multicolumn{1}{c|}{125.26}          & 56.02               \\ \hline
\textbf{office\_stringsearch1}                            & 1.28                              & \multicolumn{1}{c|}{1.80}            & 1.70                & \multicolumn{1}{c|}{6.53}            & 6.23                & \multicolumn{1}{c|}{10.58}           & 11.15               & \multicolumn{1}{c|}{40.99}           & 14.89               \\ \hline
\textbf{security\_blowfish\_d}                            & 1.03                              & \multicolumn{1}{c|}{3.69}            & 3.49                & \multicolumn{1}{c|}{13.04}           & 12.28               & \multicolumn{1}{c|}{20.82}           & 21.72               & \multicolumn{1}{c|}{75.42}           & 27.98               \\ \hline
\textbf{security\_blowfish\_e}                            & 1.14                              & \multicolumn{1}{c|}{3.55}            & 3.55                & \multicolumn{1}{c|}{12.99}           & 12.15               & \multicolumn{1}{c|}{21.57}           & 21.43               & \multicolumn{1}{c|}{75.16}           & 28.76               \\ \hline
\textbf{security\_pgp\_d}                                 & 9.45                             & \multicolumn{1}{c|}{43.40}           & 42.01               & \multicolumn{1}{c|}{173.08}          & 166.58              & \multicolumn{1}{c|}{285.13}          & 299.37              & \multicolumn{1}{c|}{1028.57}         & 403.35              \\ \hline
\textbf{security\_pgp\_e}                                 & 9.10                              & \multicolumn{1}{c|}{42.68}           & 45.05               & \multicolumn{1}{c|}{174.22}          & 167.36              & \multicolumn{1}{c|}{295.92}          & 309.43              & \multicolumn{1}{c|}{1044.20}         & 407.84              \\ \hline
\textbf{security\_sha}                                    & 1.01                             & \multicolumn{1}{c|}{2.01}            & 1.95                & \multicolumn{1}{c|}{8.03}            & 6.77                & \multicolumn{1}{c|}{12.89}           & 12.97               & \multicolumn{1}{c|}{41.99}           & 16.99               \\ \hline
\textbf{telecom\_adpcm\_c}                                & 0.52                             & \multicolumn{1}{c|}{1.12}            & 1.06                & \multicolumn{1}{c|}{2.89}            & 2.96                & \multicolumn{1}{c|}{4.42}            & 5.08                & \multicolumn{1}{c|}{16.83}           & 6.49                \\ \hline
\textbf{telecom\_adpcm\_d}                                & 0.63                              & \multicolumn{1}{c|}{1.07}            & 0.97                & \multicolumn{1}{c|}{2.79}            & 2.74                & \multicolumn{1}{c|}{4.22}            & 5.03                & \multicolumn{1}{c|}{16.12}           & 6.42                \\ \hline
\textbf{telecom\_CRC32}                                   & 0.62                              & \multicolumn{1}{c|}{0.87}            & 0.83                & \multicolumn{1}{c|}{1.95}            & 2.35                & \multicolumn{1}{c|}{3.29}            & 3.94                & \multicolumn{1}{c|}{12.17}           & 5.07                \\ \hline
\textbf{Average}               &{\textbf{7.23}} & \textbf{30.91} & \textbf{31.53} & \textbf{132.46} & \textbf{129.27} & \textbf{219.83} & \textbf{240.69} & \textbf{732.55} & \textbf{320.44}\tablefootnote{\taco{This is the direct outcome of our early exit condition which is currently set at 100.}} \\ \cline{1-10}
\end{tabular}
\caption{CBench Single-Core Build Time Analysis (Seconds) }
\label{tab:cbench_build}
\end{table*}

\taco{To showcase Protean's support for parallel builds, we modified CBench's original Makefiles so they support GNU make's \texttt{-j} option. Similar to the single core analysis, CBench does not have LTO enabled and thus the majority of the compilation time is spent during opt (for \texttt{-O3}) and protean tool (for us). We tested \texttt{-O3}, Protean compiler with IR2VEC and PFS at two settings, i.e., 50 and 500 iterations, respectively. Similar to the single-core scenario, GA initiates at the 20th iteration and guides SA during the generation of recipes. Note that both single and parallel compilation with the same setting produce the identical binaries, with all SA steps being identical to one another, and thus, due to the early stop condition being set to 100, PFS  converges earlier. Table \ref{tab:cbench_build_parallel} presents the build-time overhead of Protean when it is used in a parallel compilation setting, and the final 2 columns report the specific iterations at which both IR2VEC and PFS are early stopped. This can be used to reason about the single-core build analysis (at 500 iteration setting) reported at Table \ref{tab:cbench_build} as well. }

\begin{table*}[!t]
\centering
\small
\begin{tabular}{|l|>{\centering\arraybackslash}m{.6cm}?cc?cc?cc|}
\hline
\multirow{2}{*}{\textbf{Benchmark}} & \multirow{2}{*}{\textbf{-O3}} & \multicolumn{2}{c|}{\textbf{50 Iterations}}                 & \multicolumn{2}{c?}{\textbf{500 Iterations}}                & \multicolumn{2}{c|}{\textbf{Iterations To Converge}} \\ \cline{3-8} 
                                    &                                         & \multicolumn{1}{c|}{\textbf{IR2VEC}} & \textbf{PFS} & \multicolumn{1}{c|}{\textbf{IR2VEC}} & \textbf{PFS} & \multicolumn{1}{c|}{\textbf{IR2VEC}}    & \textbf{PFS}    \\ \hline
\textbf{automotive\_bitcount}       & 0.34                                    & \multicolumn{1}{c|}{2.67}            & 2.62                & \multicolumn{1}{c|}{15.19}           & 7.72                & \multicolumn{1}{c|}{265}                & 135                    \\ \hline
\textbf{automotive\_susan\_c}       & 5.19                                    & \multicolumn{1}{c|}{44.58}           & 43.68               & \multicolumn{1}{c|}{253.19}          & 150.75              & \multicolumn{1}{c|}{323}                & 192                    \\ \hline
\textbf{automotive\_susan\_e}       & 5.34                                    & \multicolumn{1}{c|}{44.56}           & 43.67               & \multicolumn{1}{c|}{253.08}          & 145.37              & \multicolumn{1}{c|}{333}                & 191                    \\ \hline
\textbf{automotive\_susan\_s}       & 5.10                                    & \multicolumn{1}{c|}{44.45}           & 43.56               & \multicolumn{1}{c|}{252.48}          & 139.45              & \multicolumn{1}{c|}{329}                & 182                    \\ \hline
\textbf{bzip2d}                     & 5.04                                    & \multicolumn{1}{c|}{130.40}          & 127.79              & \multicolumn{1}{c|}{740.68}          & 266.71              & \multicolumn{1}{c|}{450}                & 162                    \\ \hline
\textbf{bzip2e}                     & 4.67                                    & \multicolumn{1}{c|}{130.63}          & 143.70              & \multicolumn{1}{c|}{741.99}          & 287.70              & \multicolumn{1}{c|}{480}                & 186                    \\ \hline
\textbf{consumer\_jpeg\_c}          & 3.71                                    & \multicolumn{1}{c|}{24.28}           & 23.80               & \multicolumn{1}{c|}{137.94}          & 57.60               & \multicolumn{1}{c|}{444}                & 186                    \\ \hline
\textbf{consumer\_jpeg\_d}          & 3.58                                    & \multicolumn{1}{c|}{25.55}           & 25.04               & \multicolumn{1}{c|}{145.15}          & 62.14               & \multicolumn{1}{c|}{422}                & 181                    \\ \hline
\textbf{consumer\_lame}             & 3.90                                    & \multicolumn{1}{c|}{28.85}           & 28.28               & \multicolumn{1}{c|}{163.88}          & 66.52               & \multicolumn{1}{c|}{425}                & 172                    \\ \hline
\textbf{consumer\_mad}              & 3.24                                    & \multicolumn{1}{c|}{7.24}            & 7.09                & \multicolumn{1}{c|}{41.11}           & 15.45               & \multicolumn{1}{c|}{448}                & 168                    \\ \hline
\textbf{consumer\_tiffdither}       & 4.71                                    & \multicolumn{1}{c|}{171.66}          & 168.22              & \multicolumn{1}{c|}{975.00}          & 353.98              & \multicolumn{1}{c|}{500}                & 182                    \\ \hline
\textbf{consumer\_tiffmedian}       & 4.26                                    & \multicolumn{1}{c|}{169.92}          & 166.52              & \multicolumn{1}{c|}{965.14}          & 371.23              & \multicolumn{1}{c|}{468}                & 180                    \\ \hline
\textbf{network\_dijkstra}          & 4.39                                    & \multicolumn{1}{c|}{3.11}            & 3.05                & \multicolumn{1}{c|}{17.68}           & 8.51                & \multicolumn{1}{c|}{229}                & 110                    \\ \hline
\textbf{network\_patricia}          & 3.09                                    & \multicolumn{1}{c|}{3.31}            & 3.24                & \multicolumn{1}{c|}{18.81}           & 6.82                & \multicolumn{1}{c|}{318}                & 115                    \\ \hline
\textbf{office\_ghostscript}        & 5.50                                    & \multicolumn{1}{c|}{99.58}           & 94.60               & \multicolumn{1}{c|}{565.59}          & 251.87              & \multicolumn{1}{c|}{500}                & 275                    \\ \hline
\textbf{office\_rsynth}             & 2.69                                    & \multicolumn{1}{c|}{0.86}            & 0.84                & \multicolumn{1}{c|}{4.87}            & 2.26                & \multicolumn{1}{c|}{353}                & 164                    \\ \hline
\textbf{office\_stringsearch1}      & 3.26                                    & \multicolumn{1}{c|}{3.55}            & 3.48                & \multicolumn{1}{c|}{20.15}           & 7.73                & \multicolumn{1}{c|}{430}                & 165                    \\ \hline
\textbf{security\_blowfish\_d}      & 3.13                                    & \multicolumn{1}{c|}{7.89}            & 7.74                & \multicolumn{1}{c|}{44.84}           & 17.55               & \multicolumn{1}{c|}{384}                & 150                    \\ \hline
\textbf{security\_blowfish\_e}      & 3.02                                    & \multicolumn{1}{c|}{6.81}            & 6.67                & \multicolumn{1}{c|}{38.67}           & 14.76               & \multicolumn{1}{c|}{398}                & 152                    \\ \hline
\textbf{security\_pgp\_d}           & 3.49                                    & \multicolumn{1}{c|}{27.27}           & 26.73               & \multicolumn{1}{c|}{154.92}          & 62.80               & \multicolumn{1}{c|}{449}                & 182                    \\ \hline
\textbf{security\_pgp\_e}           & 3.75                                    & \multicolumn{1}{c|}{24.10}           & 23.61               & \multicolumn{1}{c|}{136.86}          & 50.43               & \multicolumn{1}{c|}{464}                & 171                    \\ \hline
\textbf{security\_sha}              & 3.11                                    & \multicolumn{1}{c|}{8.27}            & 8.11                & \multicolumn{1}{c|}{46.98}           & 19.63               & \multicolumn{1}{c|}{393}                & 164                    \\ \hline
\textbf{telecom\_adpcm\_c}          & 3.25                                    & \multicolumn{1}{c|}{2.67}            & 2.62                & \multicolumn{1}{c|}{15.17}           & 6.23                & \multicolumn{1}{c|}{274}                & 113                    \\ \hline
\textbf{telecom\_adpcm\_d}          & 3.52                                    & \multicolumn{1}{c|}{2.70}            & 2.65                & \multicolumn{1}{c|}{15.34}           & 6.86                & \multicolumn{1}{c|}{275}                & 123                    \\ \hline
\textbf{telecom\_CRC32}             & 3.65                                    & \multicolumn{1}{c|}{2.20}            & 2.16                & \multicolumn{1}{c|}{12.51}           & 5.52                & \multicolumn{1}{c|}{256}                & 113                    \\ \hline
\textbf{Average}                    & \textbf{3.80}                           & \multicolumn{1}{c}{\textbf{40.68}}  & \textbf{40.38}      & \multicolumn{1}{c}{\textbf{231.09}} & \textbf{95.42}      & \multicolumn{1}{c}{\textbf{310}}       & \textbf{145}           \\ \hline
\end{tabular}
\caption{CBench Parallel Build Time Analysis (64 Cores)}
\label{tab:cbench_build_parallel}
\end{table*}

\subsubsection{SPEC CPU 2017}
\label{sec:results:buildtime:spec}

\taco{Table \ref{tab:spec_build_parallel}  presents SPEC CPU 2017 single-core and parallel compilation results. Similar to the performance results reported in Table \ref{tab:spec_perf}, we run Protean compiler with IR2VEC for 30 iterations (GA initiates at the 20th iteration) and we report the build time for both \texttt{-O3} and Protean in a single-core and parallel build scenario. This setting can be controlled with \texttt{build\_ncpus=\{1, 64\}} together with \texttt{runcpu} tool. One interesting observation is that, since the baseline configuration file for SPEC CPU 2017 enables Full LTO, depending on the size of each benchmark, this unparallelizable (sequential) task can dominate the total build time for SPEC benchmarks, i.e., 500.perlbench, 502-gcc, 523-xalancbmk, etc. Therefore, leveraging multiple cores benefits Protean's build times at a higher rate than -O3s since the 30 iterations of SA happen at IR, and that is the build-time bottleneck in a single-core scenario.}

\begin{table}[!t]
\small
\begin{tabular}{|l|cc|cc|}
\hline
\multicolumn{1}{|c?}{\multirow{2}{*}{\textbf{Benchmark}}} & \multicolumn{2}{c|}{\textbf{-O3 Build Time (s)}}                                                  & \multicolumn{2}{c|}{\textbf{Protean Build Time -- 30 Iterations (s)}}                                               \\ \cline{2-5} 
\multicolumn{1}{|c?}{}                                    & \multicolumn{1}{c|}{\textbf{Single Core}} & \multicolumn{1}{c|}{\textbf{Parallel    (64 Cores)}} & \multicolumn{1}{c|}{\textbf{Single   Core}} & \multicolumn{1}{c|}{\textbf{Parallel    (64 Cores)}} \\ \hline
\textbf{500.perlbench}                                 & \multicolumn{1}{c|}{610}                  & \multicolumn{1}{c|}{420}                             & \multicolumn{1}{c|}{4800}                   & \multicolumn{1}{c|}{2760}                            \\ \hline
\textbf{502.gcc}                                       & \multicolumn{1}{c|}{2145}                 & \multicolumn{1}{c|}{1545}                            & \multicolumn{1}{c|}{8257}                   & \multicolumn{1}{c|}{5104}                            \\ \hline
\textbf{505.mcf}                                       & \multicolumn{1}{c|}{11}                   & \multicolumn{1}{c|}{6}                               & \multicolumn{1}{c|}{22}                     & \multicolumn{1}{c|}{8}                               \\ \hline
\textbf{520.omnetpp}                                   & \multicolumn{1}{c|}{130}                  & \multicolumn{1}{c|}{101}                             & \multicolumn{1}{c|}{235}                    & \multicolumn{1}{c|}{127}                             \\ \hline
\textbf{523.xalancbmk}                                 & \multicolumn{1}{c|}{601}                  & \multicolumn{1}{c|}{183}                             & \multicolumn{1}{c|}{3757}                   & \multicolumn{1}{c|}{244}                             \\ \hline
\textbf{525.x264}                                      & \multicolumn{1}{c|}{210}                  & \multicolumn{1}{c|}{131}                             & \multicolumn{1}{c|}{810}                    & \multicolumn{1}{c|}{432}                             \\ \hline
\textbf{531.deepsjeng}                                 & \multicolumn{1}{c|}{21}                   & \multicolumn{1}{c|}{14}                              & \multicolumn{1}{c|}{95}                     & \multicolumn{1}{c|}{41}                              \\ \hline
\textbf{541.leela}                                     & \multicolumn{1}{c|}{44}                   & \multicolumn{1}{c|}{25}                              & \multicolumn{1}{c|}{150}                    & \multicolumn{1}{c|}{43}                              \\ \hline
\textbf{557.xz}                                        & \multicolumn{1}{c|}{19}                   & \multicolumn{1}{c|}{11}                              & \multicolumn{1}{c|}{201}                    & \multicolumn{1}{c|}{89}                              \\ \hline
\textbf{Average}                                          & \textbf{421}                              & \textbf{271}                                         & \textbf{2036}                               & \textbf{983}                                         \\ \cline{1-5}
\end{tabular}
\caption{SPEC CPU 2017 Single Core \& Parallel (64 Cores) Build Time Analysis}
\label{tab:spec_build_parallel}
\end{table}

\section{Comparative Results}
\label{sec:comparison}

\hspace*{-1em}
\begin{table*}[!t]
\small
\centering
\begin{tabular}{|m{3cm}|cccc|cc|}
\hline
\multicolumn{1}{|c|}{\multirow{2}{*}{\textbf{Application}}}              & \multicolumn{4}{c|}{\textbf{MiCOMP's   Speedup w.r.t. -O3}}              & \multicolumn{2}{c|}{\textbf{CBench Stats}}                                \\ \cline{2-7} 
\multicolumn{1}{|c|}{}                                                   & \multicolumn{1}{c|}{10 Iter.}    & \multicolumn{1}{c|}{50 Iter.}    & \multicolumn{1}{c|}{100 Iter.}   & 500 Iter.   & \multicolumn{1}{c|}{Modules} & LOC  \\ \hline
automotive\_bitcount                                                     & \multicolumn{1}{c|}{1.001}            & \multicolumn{1}{c|}{1.013}            & \multicolumn{1}{c|}{1.028}            &  1.028            & \multicolumn{1}{c|}{10}             & 681                                                \\ \hline
automotive\_susan\_c                                                     & \multicolumn{1}{c|}{1.006}            & \multicolumn{1}{c|}{1.017}            & \multicolumn{1}{c|}{1.027}            & 1.027            & \multicolumn{1}{c|}{2}              & 2129                                               \\ \hline
automotive\_susan\_e                                                     & \multicolumn{1}{c|}{1.012}            & \multicolumn{1}{c|}{1.045}            & \multicolumn{1}{c|}{1.064}            & 1.064            & \multicolumn{1}{c|}{2}              & 2129                                               \\ \hline
automotive\_susan\_s                                                     & \multicolumn{1}{c|}{1.009}            & \multicolumn{1}{c|}{1.009}            & \multicolumn{1}{c|}{1.009}            & 1.009            & \multicolumn{1}{c|}{2}              & 2129                                              \\ \hline
bzip2d                                                                   & \multicolumn{1}{c|}{1.039}            & \multicolumn{1}{c|}{1.039}            & \multicolumn{1}{c|}{1.049}            & 1.049            & \multicolumn{1}{c|}{9}              & 6415                                               \\ \hline
bzip2e                                                                   & \multicolumn{1}{c|}{1.000}            & \multicolumn{1}{c|}{1.002}            & \multicolumn{1}{c|}{1.002}            & 1.002            & \multicolumn{1}{c|}{9}              & 6415                                               \\ \hline
consumer\_jpeg\_c                                                        & \multicolumn{1}{c|}{0.988}            & \multicolumn{1}{c|}{1.005}            & \multicolumn{1}{c|}{1.005}            & 1.005            & \multicolumn{1}{c|}{55}             & 23812                                                \\ \hline
consumer\_jpeg\_d                                                        & \multicolumn{1}{c|}{1.014}            & \multicolumn{1}{c|}{1.032}            & \multicolumn{1}{c|}{1.032}            & 1.032            & \multicolumn{1}{c|}{55}             & 23812                                              \\ \hline
consumer\_lame                                                           & \multicolumn{1}{c|}{1.017}            & \multicolumn{1}{c|}{1.017}            & \multicolumn{1}{c|}{1.017}            & 1.017            & \multicolumn{1}{c|}{35}             & 19399                                              \\ \hline
consumer\_mad                                                            & \multicolumn{1}{c|}{1.000}            & \multicolumn{1}{c|}{1.020}            & \multicolumn{1}{c|}{1.020}            & 1.020            & \multicolumn{1}{c|}{43}             & 20943                                             \\ \hline
consumer\_tiffdither                                                     & \multicolumn{1}{c|}{1.003}            & \multicolumn{1}{c|}{1.003}            & \multicolumn{1}{c|}{1.003}            & 1.003            & \multicolumn{1}{c|}{35}             & 19449                                             \\ \hline
consumer\_tiffmedian                                                     & \multicolumn{1}{c|}{1.013}            & \multicolumn{1}{c|}{1.013}            & \multicolumn{1}{c|}{1.013}            & 1.013            & \multicolumn{1}{c|}{35}             & 20021                                              \\ \hline
network\_dijkstra                                                        & \multicolumn{1}{c|}{1.035}            & \multicolumn{1}{c|}{1.075}            & \multicolumn{1}{c|}{1.075}            & 1.079            & \multicolumn{1}{c|}{1}              & 199                                               \\ \hline
network\_patricia                                                        & \multicolumn{1}{c|}{1.008}            & \multicolumn{1}{c|}{1.008}            & \multicolumn{1}{c|}{1.008}            & 1.008            & \multicolumn{1}{c|}{3}              & 577                                                \\ \hline
office\_ghostscript                                                      & \multicolumn{1}{c|}{0.980}            & \multicolumn{1}{c|}{1.010}            & \multicolumn{1}{c|}{1.010}            & 1.010            & \multicolumn{1}{c|}{288}            & 118027                                            \\ \hline
office\_rsynth                                                           & \multicolumn{1}{c|}{1.030}            & \multicolumn{1}{c|}{1.050}            & \multicolumn{1}{c|}{1.060}            & 1.060            & \multicolumn{1}{c|}{19}             & 4973                                              \\ \hline
office\_stringsearch1                                                    & \multicolumn{1}{c|}{1.002}            & \multicolumn{1}{c|}{1.004}            & \multicolumn{1}{c|}{1.004}            & 1.004            & \multicolumn{1}{c|}{4}              & 491                                               \\ \hline
security\_blowfish\_d                                                    & \multicolumn{1}{c|}{1.004}            & \multicolumn{1}{c|}{1.004}            & \multicolumn{1}{c|}{1.009}            & 1.009            & \multicolumn{1}{c|}{7}              & 870                                                \\ \hline
security\_blowfish\_e                                                    & \multicolumn{1}{c|}{0.996}            & \multicolumn{1}{c|}{1.036}            & \multicolumn{1}{c|}{1.036}            & 1.036            & \multicolumn{1}{c|}{7}              & 870                                               \\ \hline
security\_pgp\_d                                                         & \multicolumn{1}{c|}{1.031}            & \multicolumn{1}{c|}{1.044}            & \multicolumn{1}{c|}{1.057}            & 1.087            & \multicolumn{1}{c|}{39}             & 30060                                             \\ \hline
security\_pgp\_e                                                         & \multicolumn{1}{c|}{1.037}            & \multicolumn{1}{c|}{1.057}            & \multicolumn{1}{c|}{1.087}            & 1.107            & \multicolumn{1}{c|}{39}             & 30060                                           \\ \hline
security\_sha                                                            & \multicolumn{1}{c|}{1.022}            & \multicolumn{1}{c|}{1.029}            & \multicolumn{1}{c|}{1.056}            & 1.065            & \multicolumn{1}{c|}{3}              & 269                                               \\ \hline
telecom\_adpcm\_c                                                        & \multicolumn{1}{c|}{0.995}            & \multicolumn{1}{c|}{1.001}            & \multicolumn{1}{c|}{1.001}            & 1.001            & \multicolumn{1}{c|}{2}              & 305                                                \\ \hline
telecom\_adpcm\_d                                                       & \multicolumn{1}{c|}{1.001}            & \multicolumn{1}{c|}{1.040}            & \multicolumn{1}{c|}{1.040}            & 1.050            & \multicolumn{1}{c|}{2}              & 307                                                 \\ \hline
telecom\_CRC32                                                           & \multicolumn{1}{c|}{0.997}            & \multicolumn{1}{c|}{1.001}            & \multicolumn{1}{c|}{1.001}            & 1.001            & \multicolumn{1}{c|}{2}              & 213                                               \\ \hline
\multicolumn{1}{|>{\centering\arraybackslash}m{3cm}|}{\textbf{MiCOMP Speedup w.r.t. -O3 (Geomean)}}                                                         & \multicolumn{1}{c|}{\textbf{1.009}}   & \multicolumn{1}{c|}{\textbf{1.023}}   & \multicolumn{1}{c|}{\textbf{1.028}}   & \textbf{1.031}   &                  \multicolumn{2}{l}{}             \\  \cline{1-5} 
\multicolumn{1}{|>{\centering\arraybackslash}m{3cm}|}{\textbf{\taco{Protean Compiler Speedup w.r.t. -O3 (Geomean)}}} & \multicolumn{1}{c|}{\textbf{1.025}} & \multicolumn{1}{c|}{\textbf{1.035}} & \multicolumn{1}{c|}{\textbf{1.039}} & \textbf{1.041}   & \multicolumn{2}{l}{}        \\  \cline{1-5} 
\end{tabular}
\caption{Protean Compiler Performance (w/ PFS) vs. MiCOMP}
\label{tab:micomp}
\end{table*}

This section compares the performance of the Protean compiler to that of a state-of-the-art approach called MiCOMP \cite{Ashouri2017micomp}. As mentioned earlier in Section \ref{sec:related}, MiCOMP uses a Pythonic wrapper to communicate with LLVM and provide phase-ordering optimization as an iterative methodology via an ML tool called Weka \cite{weka2009}. In order to have a fair comparison, we need to take into account the following details taken from the work. First, MiCOMP's subsequences were designed with LLVM v3.8, whereas our work uses LLVM 19; this means many passes used in MiCOMP are either removed or revamped by the recent LLVM, so it is understandable that its standard baseline has also totally changed. Second, MiCOMP's feature collection method comes from a Pin tool plugin called MICA \cite{hoste2007microarchitecture}, and unfortunately, there is no aarch64 and recent Unix support on this project. \taco{For these tooling limitations, we decided to showcase the comparison between the two works by depicting the performance gains of Protean when a subsequence was found to be the optimal by the agile optimizer in a module-level scope rather than the program-wide scope, as this shows the actual performance increase that can be gained with our framework, as it enables this finer scope compared with MiCOMP's program-level limit. Therefore, we use this proxy method to bypass the two aforementioned limitations in the reproduction of the work.}  Table \ref{tab:micomp} showcases the comparison. We observe that Protean's framework has 1.6\%, 2\%, 1.4\%, and 1.1\% speedup versus MiCOMP on 10, 50, 100, and 500 iterations, respectively. In the right three columns, we also show the stats of each CBench test case and its Lines of Code (LOC), and the corresponding speedup value Protean compiler has on the 10-iteration scenario. It can be seen that as the number of modules/LOC increases, there is a higher possibility of finding a program-wide subsequence that is not the optimal case for all of its modules, and thus, the Protean compiler, with its finer-grain scope of optimization, wins against MiCOMP. 

\taco{Since MiCOMP uses MICA, each prediction has an added overhead of running the application for feature collection. Additionally, MiCOMP's recommender system and ML inference are implemented using third-party Python tools and incur disk I/O and interprocess communication overhead in its Pythonic wrapper. We estimate MiCOMP's 1-shot (Top-1 prediction) to have an average overhead of 40-45 seconds on CBench, which roughly matches Protean's 10-iteration (with PFS) overhead on its single-core compilation approach. Therefore, Protean can be seen to have an order of magnitude lower compile overhead versus MiCOMP  at an equal number of iterations on CBench. Indeed, this value can increase if we experimentally compare Protean against MiCOMP on larger benchmarks, i.e., SPEC, as their execution time will dramatically increase the feature collection round. }

\section{Integration Capabilities}
\label{sec:integration}
 
The Protean compiler is designed to be a scalable, extendable, and reproducible framework for easy integration with other third-party projects to deliver specialized benefits, and we  have already provided an integration capability by leveraging IR2VEC for feature collections. However, as discussed earlier, Protean derives an ML-based phase-ordering compilation to construct recipes of subsequences given a fine/coarse-grain code segment. This action does not alter individual optimization pass parameters, an orthogonal problem known as optimization parameter selection \cite{ashouri2016Cobayn,cummins2021compilergym}. \taco{Therefore, a couple of interesting use cases would be the following:}

\begin{enumerate}
    \item \taco{Integrate a third-party ML framework within Protean Compiler to guide parameter selection for a single pass, i.e., Function Inlining.}
    \item \taco{Integrate an LLM-based code generation strategy to guide the parameter selection together with Protean's phase-ordering capabilities.}
\end{enumerate}

\begin{figure}[!t]
\centering
\includegraphics[width=.7\textwidth]{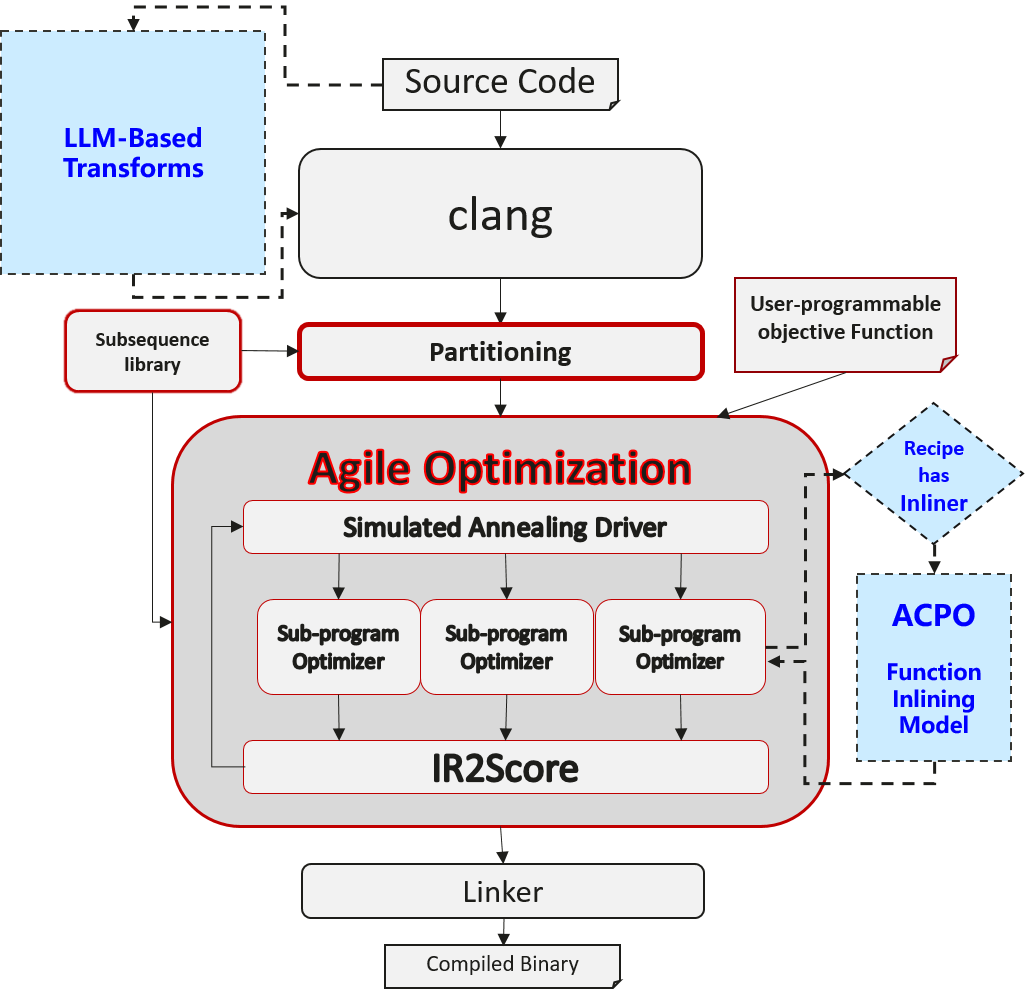}
\caption{Protean Compiler Integration Examples} 
\label{fig:protean_usecase}
\end{figure}

Note that this is by no means an exhaustive list of possibilities our framework provides, but rather an instance of multi-dimensional ML-based integration one should hope to be realized when using such frameworks. Figure \ref{fig:protean_usecase} depicts the two examples with added blue diagrams on the top left and the right-hand side of the original flow we presented earlier in Figure \ref{fig:protean_system_diagram}. \taco{During both scenarios, we run Protean with 20 iterations, and we use IR2Score trained with IR2VEC feature collection method.}

\subsection{Parameter Tuning with Third-party ML Frameworks}
\label{sec:acpo}

Protean compiler enables a fine-grain phase-ordering of optimization passes defined in the IR. Once a pass is part of the selected subsequence, a.k.a recipe, it is enabled, and the internal parameters are set by the compiler's default profitability heuristics and other checks at O3 level. Identifying the right set of optimization parameters can be defined as an extended version of the phase ordering or as an orthogonal problem \cite{ashouri2018survey}. An approach tackling the problem can pair Protean with MLGO \cite{trofin2021mlgo}, MLGOPerf \cite{ashouri2022mlgoperf}, or ACPO \cite{ashouri2023acpo,ashouri2024work}, meaning that once a recipe of subsequences is selected by Protean, the internal parameters of a single/multiple passes inside a recipe are assigned by means of leveraging the third-party ML-based frameworks. We choose the ACPO framework for this experiment, and its Function Inlining model \--- once Function Inlining is enabled as a result of choosing a recipe that has Inliner in it, i.e., \texttt{A}, the decision on whether or not to inline a callsite is granted to ACPO's Function Inlining model rather than \texttt{-O3}'s Inliner profitability heuristics. 

In order to showcase this use case, we reproduced ACPO and experimentally  tested the two-level approach on one of CBench's applications, namely, \taco{\texttt{consumer\_jpeg\_c}}.  
According to the Table \ref{tab:cbench_perf}, \texttt{-O3}'s run time on \taco{\texttt{consumer\_jpeg\_c}} is 5.84 seconds. We run Protean with 20 iterations on this benchmark using IR2VEC as its feature collection method, and we observe a 2.5\% speedup by forming the recipe \texttt{ACDCD}. Note that according to Table \ref{tab:subsequences}, subsequence A has Inliner optimization in it; therefore, by integrating ACPO's Function Inlining model with Protean, it now guides the compiler on the inlining decisions made for all 55 modules inside the benchmark, when recipe ACDCD was selected by Protean. As a result, we now observe a run time of 5.38 seconds or an additional 6\% performance speed up on top of what Protean had achieved earlier on its own, bringing the total speedup to 8.5\% over \texttt{-O3}'s performance. Upon further inspection of the decisions made by the ACPO model in the largest module of the benchmark, \texttt{jquqnt2.c}, we see \texttt{-O3}'s inliner avoids inlining callsites inside the hot functions \texttt{jinit\_2pass\_quantizer} and \texttt{fill\_inverse\_cmap}, whereas the ACPO model decided to inline and, as a result, function calls are optimized during the runs. It is worth mentioning that an 8.5\% increase in the performance comes at the cost of a 12\% binary code size increase compared to \texttt{-O3}, and a 10\% increase over Protean's initial results. \taco{Table \ref{tab:protean-acpo} depicts the detailed quantitative breakdown of this approach per module. First two columns are the list of the modules and the size, followed by the best recipe found by running Protean compiler, and if the best found recipe had function inlining optimization in it, if so, what were the percentage of the calls inlined and the generated object size for both our method and LLVM's \texttt{-O3}}

\begin{table}[!t]
\scriptsize
\centering
\begin{tabular}{|l>{\centering\arraybackslash}m{1cm}?cc>{\centering\arraybackslash}m{1.5cm}>{\centering\arraybackslash}m{1.5cm}?>{\centering\arraybackslash}m{1cm}>{\centering\arraybackslash}m{1.5cm}|}
\hline
\multicolumn{2}{|c|}{\multirow{2}{*}{\small\textbf{Consumer\_jpeg\_c}}} & \multicolumn{4}{|A{8.5cm}|}{\small\textbf{Protean Compiler (20 iterations w/ IR2VEC) + ACPO Function Inlining (FI) Model}} & \multicolumn{2}{|c|}{\multirow{2}{*}{\small\textbf{-O3}}}            \\ \hline
\textbf{Module}                  & \textbf{Module Size}            & \textbf{Best Found Recipe}         & \textbf{Object Size}        & \textbf{ACPO FI Model Used?}        & \textbf{Inlining Percentage}       & \textbf{Object Size} &  \textbf{Inlining Percentage} \\ \hline
cdjpeg.c                & 4625                   & ABCDE                     & 1808               & Yes                        & 100.0\%                    & 1808        & 100.0\%             \\ \hline
cjpeg.c                 & 19841                  & ABCDE                     & 20224              & Yes                        & 2.2\%                      & 17424       & 2.0\%               \\ \hline
jcapimin.c              & 7392                   & CD                        & 3184               & No                         & 0.0\%                      & 3448        & 0.0\%               \\ \hline
jcapistd.c              & 5881                   & ABCDE                     & 2408               & Yes                        & 0.0\%                      & 2408        & 0.0\%               \\ \hline
jccoefct.c              & 16379                  & CEAE                      & 4848               & Yes                        & 41.1\%                     & 5616        & 42.9\%              \\ \hline
jccolor.c               & 14848                  & CEAE                      & 15712              & Yes                        & 0.0\%                      & 7392        & 0.0\%               \\ \hline
jcdctmgr.c              & 12469                  & CEAE                      & 6968               & Yes                        & 0.0\%                      & 6584        & 0.0\%               \\ \hline
jchuff.c                & 25269                  & BABDA                     & 9656               & Yes                        & 83.8\%                     & 9296        & 76.2\%              \\ \hline
jcinit.c                & 2342                   & ABCDE                     & 2144               & Yes                        & 0.0\%                      & 2144        & 0.0\%               \\ \hline
jcmainct.c              & 9360                   & EECAB                     & 2192               & Yes                        & 0.0\%                      & 2192        & 0.0\%               \\ \hline
jcmarker.c              & 16647                  & CD                        & 15904              & No                         & 0.0\%                      & 13304       & 43.5\%              \\ \hline
jcmaster.c              & 19371                  & CCBDE                     & 6400               & No                         & 0.0\%                      & 8752        & 31.3\%              \\ \hline
jcomapi.c               & 2746                   & ABCDE                     & 1608               & Yes                        & 0.0\%                      & 1608        & 0.0\%               \\ \hline
jcparam.c               & 19731                  & CD                        & 9600               & No                         & 0.0\%                      & 28376       & 82.6\%              \\ \hline
jcphuff.c               & 24985                  & CD                        & 15688              & No                         & 0.0\%                      & 17360       & 26.6\%              \\ \hline
jcprepct.c              & 12073                  & ABCDE                     & 4216               & Yes                        & 55.0\%                     & 4304        & 50.0\%              \\ \hline
jcsample.c              & 18859                  & EECAB                     & 5400               & Yes                        & 55.0\%                     & 8784        & 50.0\%              \\ \hline
jctrans.c               & 13109                  & CD                        & 4784               & No                         & 0.0\%                      & 5872        & 50.0\%              \\ \hline
jdapimin.c              & 12799                  & CBCCC                     & 4184               & No                         & 0.0\%                      & 4736        & 33.3\%              \\ \hline
jdapistd.c              & 9348                   & ABCDE                     & 3352               & Yes                        & 100.0\%                    & 3472        & 100.0\%             \\ \hline
jdatadst.c              & 5119                   & ABCDE                     & 2280               & Yes                        & 0.0\%                      & 2280        & 0.0\%               \\ \hline
jdatasrc.c              & 7604                   & ABCDE                     & 2344               & Yes                        & 100.0\%                    & 2528        & 100.0\%             \\ \hline
jdcoefct.c              & 25137                  & EECAB                     & 6680               & Yes                        & 62.9\%                     & 7184        & 57.1\%              \\ \hline
jdcolor.c               & 12098                  & CAAED                     & 11520              & Yes                        & 55.0\%                     & 5352        & 50.0\%              \\ \hline
jddctmgr.c              & 8293                   & ABCDE                     & 5320               & Yes                        & 0.0\%                      & 6152        & 0.0\%               \\ \hline
jdhuff.c                & 17563                  & ACDCD                     & 16304              & Yes                        & 45.3\%                     & 11352       & 41.2\%              \\ \hline
jdinput.c               & 13497                  & EECAB                     & 4312               & Yes                        & 47.1\%                     & 4184        & 42.9\%              \\ \hline
jdmainct.c              & 20645                  & ABCDE                     & 4048               & Yes                        & 55.0\%                     & 4448        & 50.0\%              \\ \hline
jdmarker.c              & 30754                  & BABDA                     & 15360              & Yes                        & 82.5\%                     & 13944       & 75.0\%              \\ \hline
jdmaster.c              & 19670                  & BABDA                     & 5896               & Yes                        & 55.0\%                     & 7352        & 50.0\%              \\ \hline
jdmerge.c               & 13916                  & EECAB                     & 6648               & Yes                        & 55.0\%                     & 4240        & 50.0\%              \\ \hline
jdphuff.c               & 19691                  & ECECB                     & 7808               & No                         & 0.0\%                      & 7896        & 100.0\%             \\ \hline
jdpostct.c              & 9723                   & ABCDE                     & 3216               & Yes                        & 0.0\%                      & 3216        & 0.0\%               \\ \hline
jdsample.c              & 16381                  & EECAB                     & 5208               & Yes                        & 0.0\%                      & 6912        & 0.0\%               \\ \hline
jdtrans.c               & 4141                   & ABCDE                     & 1952               & Yes                        & 55.0\%                     & 1992        & 50.0\%              \\ \hline
jerror.c                & 6798                   & ABCDE                     & 11384              & Yes                        & 0.0\%                      & 11384       & 0.0\%               \\ \hline
jfdctflt.c              & 5486                   & ABCDE                     & 1536               & Yes                        & 0.0\%                      & 2256        & 0.0\%               \\ \hline
jfdctfst.c              & 7578                   & CBCCC                     & 1560               & No                         & 0.0\%                      & 1888        & 0.0\%               \\ \hline
jfdctint.c              & 11066                  & BABDA                     & 1824               & Yes                        & 0.0\%                      & 1824        & 0.0\%               \\ \hline
jidctflt.c              & 8418                   & ACDCD                     & 1952               & Yes                        & 0.0\%                      & 1952        & 0.0\%               \\ \hline
jidctfst.c              & 13104                  & ACDCD                     & 1968               & Yes                        & 0.0\%                      & 1960        & 0.0\%               \\ \hline
jidctint.c              & 14749                  & CD                        & 2184               & No                         & 0.0\%                      & 2200        & 0.0\%               \\ \hline
jidctred.c              & 13442                  & CD                        & 3784               & No                         & 0.0\%                      & 3792        & 0.0\%               \\ \hline
jmemansi.c              & 4610                   & ABCDE                     & 3248               & Yes                        & 0.0\%                      & 3248        & 0.0\%               \\ \hline
jmemmgr.c               & 40884                  & CDAEC                     & 10968              & Yes                        & 40.7\%                     & 10584       & 37.0\%              \\ \hline
jquant1.c               & 31294                  & CEAE                      & 17976              & Yes                        & 58.2\%                     & 12040       & 52.9\%              \\ \hline
jquant2.c               & 48429                  & ADCCD                     & 13720              & Yes                        & 25.7\%                     & 15848       & 23.3\%              \\ \hline
jutils.c                & 5240                   & EECAB                     & 2248               & Yes                        & 0.0\%                      & 2248        & 0.0\%               \\ \hline
loop-wrap.c             & 454                    & CDAEC                     & 2104               & Yes                        & 0.0\%                      & 2104        & 0.0\%               \\ \hline
rdbmp.c                 & 13792                  & CD                        & 6632               & No                         & 0.0\%                      & 7536        & 90.0\%              \\ \hline
rdgif.c                 & 22939                  & ABCDE                     & 11328              & Yes                        & 58.7\%                     & 7760        & 53.3\%              \\ \hline
rdppm.c                 & 13731                  & ABDBB                     & 5456               & Yes                        & 24.4\%                     & 5808        & 22.2\%              \\ \hline
rdrle.c                 & 11673                  & ACDCD                     & 728                & Yes                        & 0.0\%                      & 728         & 0.0\%               \\ \hline
rdswitch.c              & 9675                   & CD                        & 7680               & No                         & 0.0\%                      & 7904        & 16.7\%              \\ \hline
rdtarga.c               & 14967                  & ACDCD                     & 6896               & Yes                        & 96.3\%                     & 6752        & 87.5\%              \\ \hline
\multicolumn{3}{|c|}{\footnotesize\textbf{Runtime Speedup (Protean+ACPO):\quad  2.5\% + 6\%}} & \textbf{354 KB} & -- & \textbf{41.5\%} & \textbf{324 KB}  & \textbf{31.4\%}  \\ \hline
\end{tabular}
\caption{Fine-grain Parameter Tuning with Protean + ACPO Function Inlining Model}
\label{tab:protean-acpo}
\end{table}

\subsection{Code Optimization with LLM Integrations}
\label{sec:llms}

To further demonstrate the utility of the Protean compiler, we evaluate its effectiveness in tandem with LLM-based source code optimizations. In this flow, a Large Language Model (LLM) is used to optimize the initial workload in a source-to-source fashion, allowing the functionality to remain the same, but the source code implementation to differ. \cc{The motivation is that sometimes the original phase ordering flow does not yield the maximum benefits due to the compiler's transformation passes being too conservative in profitability checks, a lack of heuristics to support corner cases, or simply the inability to transform the code when the pass(es) are enabled by the Protean phase ordering.}
We use Qwen2.5 - 32B - Instruct ~\cite{hui2024qwen2} model to generate optimized candidates that are then validated to ensure the transformed code is indeed producing the correct output. We use the runtime performance we obtained by compiling the LLM-optimized code as a baseline for comparison against compiling the same LLM-optimized source code with the Protean Compiler. Note that this approach, i.e., LLM-based code generation and optimization, can be further automated in various fashions and is an orthogonal research topic; for brevity, we showcase one full example of this approach with our benchmark application, and we leave the full-fledged automated framework for our future work. 

According to Table \ref{tab:micomp}, \texttt{automotive\_susan\_c} has two modules, and the bulk of code is located inside \texttt{susan.c}. This module can be optimized by our LLM by means of unrolling with a factor of 5. Therefore, we have two versions of the function, \texttt{susan\_corners()}, by which we compile both and evaluate the runtimes. We note that the LLM version now has a 3.1\% speedup with respect to LLVM's \texttt{-O3}'s. Note that according to our results shown under Table \ref{tab:cbench_perf}, this application was sped up by 3.6\% by our Protean framework's phase-ordering flow; however, this time we feed the LLM optimized version to our framework to see if there are additional speed-up values to be had. The reason is that the starting \texttt{state} of the test case has now changed, albeit the functionality is identical, and thus, the agile driver can leverage this and find a new optimal solution for it. 

\begin{figure*}[!h]
\scriptsize
\begin{lstlisting}[language=C++, label={list:llms}, caption=Code Optimization using Protean Compiler and LLM Integration]
// Iterations = 20, Temperature (Min, Max) = (0,100), Geometric cooling,
// Infererence Model: IR2Score trained w/ IR2VEC feature collection method
>  clang -OP -mllvm -protean 
   -Wprotean,-use-protean-collect=false,-max-iterations=20,-protean-output-table  
   *.c -lm 
...
ProteanCompiler :: Beginning Simulated Annealing...
--------------------------------------------------
ProteanCompiler :: Optimizing module "loop-wrap.c"
-------------------------------------------------
Iter  Current State    Next State   Best State  Current Cost   Next Cost   Best Cost   Temp
0        ABCDE           ABCDE        ABCDE        1.01          1.01         1.01     100.000
1        ABCDE           CBCCC        ABCDE        1.00          1.00         1.01      79.432
...
18       DBBBC           ABBBA        ABCDE        1.00          1.00         1.01      1.584
19       ABBBA           DCCBA        ABCDE        1.00          1.00         1.01      1.258
...
------------------------------------------------
ProteanCompiler :: Optimizing module "susan.c"
------------------------------------------------
Iter  Current State    Next State   Best State  Current Cost   Next Cost    Best Cost   Temp
0        ABCDE           ABCDE        ABCDE        1.05          1.05         1.05     100.000
1        ABCDE           CBCCC        ABCDE        1.05          1.06         1.05      79.432
2        CBCCC           BABDA        CBCCC        1.06          1.05         1.06      63.095
...
8        ECADB           CBABE        CBCCC        1.03          1.04         1.06      15.848
9        CBABE           ACDCD        CBCCC        1.04          1.01         1.06      12.589
10       ACDCD              CD        CBCCC        1.01          1.08         1.06      10.000
11          CD           CDAEC           CD        1.08          1.07         1.08       7.943
12       CDAEC           CAAED           CD        1.07          1.07         1.08       6.309
...
18       DBBBC           ABBBA           CD        1.03          1.05         1.08       1.584
19       DBBBC           DCCBA           CD        1.03          1.06         1.08       1.258

Explored Recipes Size: 20
ProteanCompiler :: Simulated Annealing finished running for Module /tmp/susan.bc
The final recipe accepted is "CD":
function<eager-inv>(sroa<modify-cfg>,function(gvn-hoist,mldst-motion,gvn,sccp,bdce,instcombine<max-iterations=1;no-use-loop-info;no-verify-fixpoint>,jump-threading,correlated-propagation,adce),memcpyopt),cgscc(dse,function<eager-inv>(loop-simplify,function(lcssa,coro-elide,simplifycfg<bonus-inst-threshold=1;no-forward-switch-cond;no-switch-range-to-icmp;no-switch-to-lookup;keep-loops;no-hoist-common-insts;no-sink-common-insts;speculate-blocks;simplify-cond-branch>,instcombine<max-iterations=1;no-use-loop-info;no-verify-fixpoint>),reassociate),cgscc(function-attrs,function(require<should-not-run-function-passes>),coro-split,function(invalidate<all>))),deadargelim,coro-cleanup,globalopt,globaldce,elim-avail-extern,rpo-function-attrs,recompute-globalsaa,ipsccp,function<eager-inv>(float2int,lower-constant-intrinsics),constmerge,cg-profile,rel-lookup-table-converter,ir-libraryinjection

\end{lstlisting}
\end{figure*}

Listing \ref{list:llms} showcases the inference flow of Protean by 20 iterations when used with the revised test case. As expected, the smaller module without a change, shown on top, doesn't have any potential opportunities and thus, the agile driver doesn't accept any better steps than the initial, and all are in fact around 1, showing no predicted module speedup by the IR2Score model. However, when optimizing the revised \texttt{susan.c} module, we quickly see the simulated annealing driver exploring locally optimal points, one after another, and as long as the temperature remains hot and the better solutions are getting accepted, until we visit recipe CD, which has the predicted module speed up of around 8\% that turns out to be the best solution found at the 11th iteration. The \textit{Best State} column shows that visiting a few other states did not yield any more optimal solutions, and thus the driver ends its run successfully by showcasing the subsequence CD and its internal optimization passes (refer to Table \ref{tab:subsequences} for the dictionary). Upon finishing the agile driver, the compiler starts linking its modules and generates the binary, and it has a 7\% speed up compared to the LLM-optimized source code. 

Note that given the small generalization (test) error we showcase in Figure \ref{fig:test}, we expect the final binary to have a similar actual speedup value to \texttt{susan.c}'s predicted module speedup of 8\% because it is by far the only contributing module of the benchmark, and secondly, standard Makefiles of CBench benchmark don't apply Link-time-optimization (LTO) to generate the binaries. Therefore, this example showcases the LLM integration capabilities of the Protean compiler, for which we managed to speed up our test case by 3.1\% using the LLM model, plus an additional 7\% using the Protean compiler, respectively.

\section{Discussion \& Conclusion}
\label{sec:conclusion}

\paragraph{Multidimensional ML Framework Extension} As discussed earlier, Protean compiler extends seamlessly to support a multitude of integration scenarios, including but not limited to leveraging third-party feature/embeddings generation (IR2VEC), an ML-based framework (ACPO), and LLM-based code optimization. The latter is an exciting area of research and development that we are actively pursuing in our future work to address several automation, verification, and reliability challenges around it. 
Additionally, we showed how multidimensional ML frameworks can be used to formulate the optimization problems at both the algorithmic level and a fine-grained level. This helps to piece together the puzzles of a selection and the phase-ordering problem of optimizations, thus enhancing the speedup values gained of up to 8.5\% and 10.1\% on two selected CBench applications, respectively. 

\paragraph{Scalable Feature Generation Library} Currently, we have implemented a complete library of handcrafted features at different scopes of a program, and we have already integrated IR2VEC into our flow. The Protean compiler can be easily extended to support other types of feature methods, i.e., graph-based, encoder-based embeddings, etc., and we plan to incrementally enhance the feature set of our existing library as well. Finally, with the fast pace of incoming LLMs, we aim to continuously release better and more accurate IR2Score models, which in turn enhance the quality of the decisions our agile compiler takes in converging to more optimal points more rapidly.

\paragraph{Code Scope Scalability} Protean compiler supports a fine-grain scope of code optimization in its agile driver, meaning that we can adapt the agile compilation to run on a program-wide, module, function, or loop-level scope of the code.  In this work and among all the possibilities, we decided to showcase the benefits at a module-level scope, and this certainly doesn't limit the scope only to that. Users might benefit from another level of granularity, and the framework and the Protean feature set allow them to build that strategy easily. 

This work proposed and showcased the Protean compiler, the first end-to-end compiler phase-ordering framework that leverages an agile driver to automatically construct optimized phase-ordering recipes of compiler optimization at the fine-grained scope of programs. Our experimental results show speedup values up to 4.1\% on average and up to 15.7\% on select CBench applications, respectively, with only a minimal build-time overhead. Additionally, \taco{Protean speeds up SPEC CPU 2017 (Integer suite) by an average of 1\% and up to 6.4\% on selected applications.} We show the integration capabilities of the Protean compiler, which gained a 10.1\% speedup value on the Susan and 8.5\% on the Jpeg test cases. We plan to release the code base to the open-source community in the near future.    


\bibliographystyle{ACM-Reference-Format}
\balance
\bibliography{acmart}

\newpage
\onecolumn
\appendix

\section{Supplementary Material}

This section provides a set of supplementary information on the Protean Compiler framework, including but not limited to the PFS at different scopes of the code, a short list of hyperparameters supported by the framework, etc.  

\subsection{Protean Feature Set (PFS)}

This is the full list of Protean Feature Set (PFS), which are shown as examples. We have generated \texttt{loop-wrap.ll} from \texttt{loop-wrap.c} under \texttt{automotive-susan-c$/$src} and run the pass to dump its features. 

As can be seen, the file is a one-function and one-loop  tiny module, and thus, the PFS will dump two rows of features containing the module level features at rows 1 and at row 2, where we have the function and the loop features exist in the IR file. In the first row, the loop and call features are zero because the level is the module level, and in the second row, we have the loop features dumped and the caller\/callee features dumped at zero. This way, we can generate a scope-aware feature set and aggregate them for larger modules, if necessary. Note that larger modules can have multiple rows generated as a result of having a higher number of loops and functions.

\begin{figure*}[!t]
\scriptsize
\begin{lstlisting}[language=C++, label={list:pfs}, caption=Protean Feature Set (PFS) Full List Dump]
//loop_wrap.c
  1 #include <stdio.h>
  2
  3 void main1(int argc, char* argv[]);
  4
  5 int main(int argc, char* argv[])
  6 {
  7   FILE* loop_wrap=NULL;
  8   long loop_wrap1, loop_wrap2;
  9
 10   if ((loop_wrap=fopen("_finfo_dataset","rt"))==NULL)
 11   {
 12     fprintf(stderr,"\nError: Can't find dataset!\n");
 13     return 1;
 14   }
 15
 16   fscanf(loop_wrap, "%ld", &loop_wrap2);
 17   fclose(loop_wrap);
 18
 19   for (loop_wrap1=0; loop_wrap1<loop_wrap2; loop_wrap1++)
 20   {
 21     main1(argc, argv);
 22   }
 23
 24   return 0;
 25 }

//> build/bin/protean -passes=protean-collect-features  --debug-only=proteanFC --enable-protean-feature-dump loop-wrap.ll
Module|Function|Callee|Caller|Loop,average-store-instructions-per-function,average-load-instructions-per-function,average-instructions-per-function,global-variable-count,critical-edge-count,total-edge-count,loop-count,median-calls-per-function,average-calls-per-function,total-function-calls,total-instruction-count,average-bb-per-function,total-bb-count,function-count,node-count,edge-count,is-tail,is-must-tail,is-in-inner-loop,is-indirect,opt-code,mandatory-only,mandatory-kind,loop-level,cost-estimate,nr-ctant-params,callsite-height,block-freq,..., TotLoopNestInstCount,NumStoreInstPerLoopNest,NumLoadInstPerLoopNest,AvgNumInsts,AvgStoreSetSize,IndVarSetSize,NumPartitions,StepValueInt,FinalIVValueInt,InitialIVValueInt,Size,MaxTripCount,TripCount

loop-wrap.ll||||,9.000000,9.000000,40.000000,5,0,9,1,0,0.000000,0,40,64.0,64,1,1, 0,0,0,0,0,0,0,0,0,0,0,0,0,0,0,0,0,0,0,0,0,0,0,0,0,0,0,0,0,0,0,0,0,0,0,0,0,0,0,0,0,0,0 ,0,0,0,0,0,0,0,0,0,0,0,0,0,0,0,0,0,0,0,0,0,0,0,0, 0,0,0,0,0,0,0,0,0,0,0,0,0,0,0,0,0,0,0,0,0,0,0,0, 0,0,0,0,0,0,0,0,0,0,0,0,0,0,0,0,0,0,0,0,0,0,0,0,0,0,0,0,0,0,0,0,0,0
loop-wrap.ll|main|||for.cond,9.000000,9.000000,40.000000,5,0,9,1,0,0.0,0,40,64.000000,64,1,1, 0,0,0,0,0,0,0,0,0,0,0,0,0,0,0,0,0,0,0,0,0,0,0,0,0,0,0,0,0,0,0,0,0,0,0,0,0,0,0,0,0, 0,0,0,0,0,0,0,0,0,0,0,0,0,0,0,0,0,0,0,0,0,0,0,0,0,0,0,0,0,0,0,0,0,0,0,0,0,0,0,0,0, 0,0,0,0,0,0,1,1,3,0.416667,12,1,5,0.416667,12,1,5,0.000000,0.000000, 0,0,0,0,0,14,0,0


\end{lstlisting}
\end{figure*}

\subsection{Protean Compiler Hyperparameters}

Here is the list of Protean framework hyperparameters defined. The default values are used in generating the result of the current version of the paper unless specified otherwise in the text.

\begin{figure*}[!t]
\scriptsize
\lstset{escapeinside=``}
\begin{lstlisting}[language=C++, label={list:hyperparam}, caption=Protean Framework Hyperparameters]
//===----------------------------------------------------------------------===/
// List of Protean Framework Hyperparameters (with their default values)

static cl::opt<CoolingType> CoolingSchedule(
    "cooling", cl::init(Geometric),
    cl::desc("Choose Cooling Schedule for Simulated Annealing"),
    cl::values(clEnumVal(Geometric, "Determine cost based on file size"),
               clEnumVal(Linear, "Determine cost based on instruction count")));
static cl::opt<unsigned> MaxIterations(
    "max-iterations",
    cl::desc("Specify Maximum Iterations for Simulated Annealing"),          
    cl::init(100));
static cl::opt<unsigned>
    RngVal("rng-val", cl::desc("Specify RNG Val for Simulated Annealing"),//Reproducible runs 
           cl::init(123));
static cl::opt<double> MaxTemperature(
    "max-temperature",
    cl::desc("Specify Maximum Temperature for Simulated Annealing"),
    cl::init(100));
static cl::opt<unsigned> InitialSampleSize(
    "initial-sample-size",
    cl::desc(
        "Specify Number of Initial Random Samples for Simulated Annealing"),  
    cl::init(20));
static cl::opt<double>
    MutationRate("mutation-rate",
                 cl::desc("Specify Mutation Rate for Genetic Recommender"),
                 cl::init(0.05));
static cl::opt<double>
    CrossoverRate("crossover-rate",
                  cl::desc("Specify Crossover Rate for Genetic Recommender"),
                  cl::init(0.95));
static cl::opt<double>
    PopulationSize("population-size",
                   cl::desc("Specify Population Size for Genetic Recommender"),
                   cl::init(10));
static cl::opt<CrossoverFunction> CrossoverType(
    "crossover-type", cl::init(SinglePoint),
    cl::desc("Choose crossover method type for Genetic Recommender"),
    cl::values(clEnumVal(SinglePoint, "Single point crossover method"),
               clEnumVal(DoublePoint, "Double point crossover method"),
               clEnumVal(Uniform, "Uniform crossover method")));
static cl::opt<MutationFunction> MutationType(
    "mutation-type", cl::init(FlipOne),
    cl::desc("Choose mutation method type for Genetic Recommender"),
    cl::values(clEnumVal(FlipOne, "Mutate one sequence mutation method"),
               clEnumVal(SwapTwo, "Swap two sequences mutation method")));
static cl::opt<IRCostFunction> CostType(
    "cost-type", cl::init(IRAnalysis),
    cl::desc("Choose IR Cost Function used for Simulated Annealing"),
    cl::values(
        clEnumVal(IRAnalysis, "Determine cost based on IR2Score"),         // IR2Score fitness
        clEnumVal(MCA, "Determine cost based on llvm-mca (cycle count)")));// llvm-mca fitness
static cl::opt<bool>
    UseProteanCollect("use-protean-collect",
                      cl::desc("Use protean collect features for IR Analyzer"), 
                      cl::init(false));
static cl::opt<bool>
    ModLevelIPC("module-level-ipc",
                cl::desc("Enable IPC for module level shared memory"), // Avoid unneccesaryI/O 
                cl::init(false));                                      // ON at feature dump
//===----------------------------------------------------------------------===//


\end{lstlisting}
\end{figure*}

\end{document}